\documentclass[twocolumn,aps,pra,notitlepage,superscriptaddress,longbibliography,nofootinbib]{revtex4-1}
\usepackage[breaklinks=true]{hyperref}
\hypersetup{
    colorlinks,
    linkcolor={red!50!black},
    citecolor={blue!50!black},
    urlcolor={blue!80!black}
}
\usepackage{amsmath,amssymb,amsthm,mathtools,mathrsfs}
\usepackage[table,dvipsnames]{xcolor}
\usepackage{tikz}
\usetikzlibrary{quantikz2}
\usepackage{adjustbox}
\usepackage{dsfont}
\usepackage[normalem]{ulem}
\usepackage{enumerate}
\usepackage[all]{xy} 

\theoremstyle{plain}
\newtheorem{theorem}{Theorem}
\newtheorem{lemma}{Lemma}

\theoremstyle{definition}
\newtheorem{definition}{Definition}

\newcommand{\bd}{\begin{definition}}
\newcommand{\ed}{\end{definition}}
\newcommand{\bt}{\begin{theorem}}
\newcommand{\et}{\end{theorem}}
\newcommand{\be}{\begin{equation}}
\newcommand{\ee}{\end{equation}}
\newcommand{\blem}{\begin{lemma}}
\newcommand{\elem}{\end{lemma}}

\newcommand\define[1]{\emph{\textbf{#1}}}

\newcommand{\matr}{\mathbb{M}}

\DeclareMathAlphabet{\mathpzc}{OT1}{pzc}{m}{it} 
 \DeclareFontFamily{OT1}{pzc}{}
 \DeclareFontShape{OT1}{pzc}{m}{it}{ <-> s*[1.2] pzcmi7t }{}
 \DeclareMathAlphabet{\mathpzc}{OT1}{pzc}{m}{it}
 \newcommand{\Alg}[1]{\mathpzc{#1}}

\newcommand{\obs}[1]{\mathscr{O}_{#1}}
\newcommand{\map}[1]{\mathcal{#1}}

\newcommand{\Ad}{\mathrm{Ad}}
\def\R{{{\mathbb R}}}
\def\C{{{\mathbb C}}}
\newcommand{\<}{\langle}
\renewcommand{\>}{\rangle}

\newcommand{\Tr}{\operatorname{Tr}}

\newcommand{\Jamiol}{\mathscr{J}}
\newcommand{\Choi}{\mathscr{C}}

\begin{document}									
\preprint{APS/123-QED}

\title{Time-symmetric correlations for open quantum systems}

\author{Arthur J. Parzygnat}
\email{arthurjp@mit.edu}
\affiliation{Experimental Study Group, Massachusetts Institute of Technology, Cambridge, Massachusetts 02139, USA.}
\author{James Fullwood}
\email{fullwood@hainanu.edu.cn}
\affiliation{School of Mathematics and Statistics, Hainan University, Haikou, Hainan Province, 570228, China.}

\date{\today}

\begin{abstract}
Two-time expectation values of sequential measurements of dichotomic observables are known to be time symmetric for closed quantum systems. Namely, if a system evolves unitarily between sequential measurements of dichotomic observables $\obs{A}$ followed by $\obs{B}$, then it necessarily follows that $\<\obs{A}\,,\obs{B}\>=\<\obs{B}\,,\obs{A}\>$, where $\<\obs{A}\,,\obs{B}\>$ is the two-time expectation value corresponding to the product of the measurement outcomes of $\obs{A}$ followed by $\obs{B}$, and $\<\obs{B}\,,\obs{A}\>$ is the two-time expectation value associated with the time reversal of the unitary dynamics, where a measurement of $\obs{B}$ precedes a measurement of $\obs{A}$. In this work, we show that a quantum Bayes' rule implies a time symmetry for two-time expectation values associated with \emph{open} quantum systems, which evolve according to a general quantum channel between measurements. Such results are in contrast with the view that processes associated with open quantum systems---which may lose information to their environment---are not reversible in any operational sense. We give an example of such time-symmetric correlations for the amplitude-damping channel, and we propose an experimental protocol for the potential verification of the theoretical predictions associated with our results.
\end{abstract}

	\maketitle

\section{Introduction}
\label{sec:intro}
Correlations between a pair of classical random variables make no distinction between space and time. For example, if $X$ and $Y$ are random variables on a population, then the joint statistics of $X$ and $Y$ will be indifferent to whether or not one measures $X$ and $Y$ successively on a single random sample, or in parallel on two distinct random samples. 
Quantum correlations, however, are sensitive to the way in which they are measured~\cite{vN18}, thus resulting in a more refined theory in which a distinction between space and time becomes manifest.

While spacelike quantum correlations are at the heart of many foundational aspects of quantum theory---such as quantum entanglement and the failure of local realism~\cite{Sc35,EPR,Bell64,CHSH69,Fine82,We89,HHHH09,NiCh11,Witten18}---quantum correlations across space \emph{and} time are much less understood~\cite{LeGa85,BMKG13,LeSp13,FJV15,ZPTGVF18,PZOVF19,PaRu19,PaBayes,ZDV20,BJDJMSX20,PaRuBayes,GPRR21,SSS20,JSK23,song_2023,LQDV23,LCD24}. For example, there is no well-established notion of entanglement for timelike-separated systems~\cite{BTCV04,OlRa11,OlRa12,Jia17,Ra20,MVVRAPGDG19,ZCPB19,MVVRAPGDG20,MVVAPGDG21,NTTTW21,DHMTT22,DHMTT23}. One reason for this is due to a disparity in the mathematical formalisms with which quantum theory describes spatial versus temporal correlations. In particular, while spatial correlations are encoded by density operators, temporal correlations are encoded by quantum channels describing the dynamical evolution of quantum systems~\cite{vN18,Kr83,NiCh11,BCS04,Pa17,FuJa13}. 

The issue of extending the density operator formalism to the temporal domain has been addressed in many works, including Refs.~\cite{Wat55,ABL64,Ja72,Ch75,Oh83a,Oh83b,ReAh95,NiSh96,Le06,Le07,AhVa08,APTV09,CDP09,APT10,OCB12,BDOV13,GJM13,LeSp13,BDOV14,ABCFGB15,FJV15,HHPBS17,CJQW18,CRGWF18,ChMa20,KhMo21,PaQPL21,MaCh22,FuPa22,FuPa22a,JiKa23,DMR24}, which has resulted in various approaches towards providing a more unified treatment of space and time in quantum theory. In this work, we make use of the spatiotemporal formalism provided by \emph{quantum states over time} (as developed in Refs.~\cite{HHPBS17,FuPa22,FuPa22a,FuPa24a,Fu23,Fu23a,PFBC23,LiNg23}) to establish an operational time symmetry for the theoretical two-time expectation values associated with sequential measurements on an open quantum system. While the uncertainty principle implies that the measurement outcomes of incompatible observables depend on the order in which they are measured, we use the quantum Bayes' rule established in Ref.~\cite{FuPa22a} to show that the expectation value of the \emph{product} of the measurement outcomes of dichotomic observables is independent of the order in which they are measured, even when the observables do not commute. Such time-symmetric correlations for open quantum systems are quite unexpected, as open quantum systems may lose information to their environment and are hence not reversible. 

In light of the operational nature of our results, we formulate concrete and testable predictions regarding time-symmetric expectation values for open quantum systems, which evolve according to a general quantum channel between measurements. We work out such predictions for the case of amplitude-damping channels, and then conclude by proposing an experimental implementation.

\section{Preliminaries}
In this work, $A$ and $B$ denote finite-dimensional quantum systems, and $\{ | i\rangle\}_{i=1}^{d}$ denotes an orthonormal basis of the Hilbert space $\mathcal{H}_A$ associated with $A$. We denote the two algebras of linear operators on $\mathcal{H}_A$ and $\mathcal{H}_B$ by $\Alg{A}$ and $\Alg{B}$ and the two spaces of observables on $A$ and $B$ by $\mathbf{Obs}(A)$ and $\mathbf{Obs}(B)$, respectively. To make our results as mathematically precise as possible, we now provide the necessary definitions and terminology for the statement of our main theorem, which appears in the next section. Further details and motivations may be found in the standard references~\cite{vN18,Ja72,Ch75,Kr83,Paulsen02,BCS04,NiCh11,BeZy06,Sakurai20}.

A \define{quantum channel} from a quantum system $A$ to a quantum system $B$ consists of a completely positive, trace-preserving  linear map $\map{E}:\Alg{A}\to \Alg{B}$. 
General quantum channels are used to model the evolution of \emph{open} quantum systems. 
This is in contrast to closed quantum systems, whose evolution is governed by unitary dynamics. Given a quantum channel $\map{E}:\Alg{A}\to \Alg{B}$, there exists a collection of \define{Kraus operators} $\{E_{\alpha}\}$ such that
$\map{E}(\rho)=\sum_{\alpha}E_{\alpha}\rho E_{\alpha}^{\dag}$ for all $\rho\in \Alg{A}$ and $\sum_{\alpha}E_{\alpha}^{\dag}E_{\alpha}=\mathds{1}_{A}$.
Such a collection of Kraus operators $\{E_{\alpha}\}$ is said to be a \define{Kraus representation} of the channel $\map{E}$.

\bd
Suppose that a system $A$, initially in the state $\rho\in \Alg{A}$, evolves according to a quantum channel $\map{E}:\Alg{A}\to \Alg{B}$ between measurements of observables $\obs{A}\in\mathbf{Obs}(A)$ and $\obs{B}\in\mathbf{Obs}(B)$. Then the \define{two-time expectation value} associated with the sequential measurement of $\obs{A}$ followed by $\mathscr{O}_{B}$ is the real number $\langle \obs{A}\,, \obs{B} \rangle$ given by
\begin{align} \label{2TEXT57}
\langle \obs{A}\,, \obs{B} \rangle=\sum_{i}\lambda_i\Tr\Big[\map{E}(P_i\rho P_i)\obs{B}\Big]\, ,
\end{align}
where $\obs{A}=\sum_{i}\lambda_i P_i$ is the spectral decomposition of $\obs{A}$. 
\ed

The expression $\langle \obs{A}\,, \obs{B} \rangle$, as given by~\eqref{2TEXT57}, is referred to as the two-time expectation value of $\obs{A}$ followed by $\obs{B}$ \emph{with respect to} the pair $(\map{E},\rho)$~\cite{FuPa24a}, the latter of which is called a \define{process}. In what follows, we define two separate notions of inverse associated with a process $(\map{E},\rho)$.

\bd
\label{defn:operationalinverse}
Let $\map{E}:\Alg{A}\to \Alg{B}$ be a quantum channel and let $\rho\in\Alg{A}$ be a state.  
Given a subset $\mathscr{S}\subset \mathbf{Obs}(A)\times \mathbf{Obs}(B)$, a quantum channel $\map{F}:\Alg{B}\to \Alg{A}$ is said to be an \define{$\mathscr{S}$-operational inverse} of the process $(\map{E},\rho)$ iff 
\be
\label{eqn:BayessymmetricTTEV}
\langle \obs{A}\,, \obs{B} \rangle=\langle \obs{B}\,, \obs{A} \rangle
\ee
for every pair $(\obs{A},\obs{B})\in \mathscr{S}$, where $\langle \obs{A}\,, \obs{B} \rangle$ and $\langle \obs{B}\,, \obs{A} \rangle$ are the two-time expectation values with respect to $(\map{E},\rho)$ and $(\map{F},\map{E}(\rho))$, respectively.
\ed

Our main theorem involves $\mathscr{S}$-operational inverses for the subset $\mathscr{S}\subset \mathbf{Obs}(A)\times \mathbf{Obs}(B)$ consisting of all pairs of \emph{light-touch observables}, which we now define.

\bd
An observable is \define{light-touch} iff its set of distinct eigenvalues is of the form $\{\pm \lambda\}$ or $\{\lambda\}$ for some $\lambda\in\R$. A light-touch observable is \define{dichotomic} iff its set of distinct eigenvalues is $\{\pm 1\}$. The set of all light-touch observables on a quantum system $X$ is denoted by $\mathscr{L}_X$. 
\ed

We now recall the definition of the Jamio{\l}kowski matrix, which we use to formulate a spatiotemporal analogue of a bipartite density matrix. 

\bd
\label{defn:Jamiolkowski}
Let $\map{E}:\Alg{A}\to \Alg{B}$ be a quantum channel. 
The \define{Jamio{\l}kowski matrix} associated with $\map{E}$ is the element $\mathscr{J}[\map{E}]\in \Alg{A}\otimes \Alg{B}$ given by
\be \label{JMX57}
\mathscr{J}[\map{E}]=\sum_{i,j}|i\rangle \langle j|\otimes \map{E}(|j\rangle \langle i|)\, .
\ee
\ed

\bd
Given a quantum channel $\map{E}:\Alg{A}\to \Alg{B}$ and a state $\rho\in \Alg{A}$, the \define{spatiotemporal product} of $\map{E}$ and $\rho$ is the element $\map{E}\star \rho$ in $\Alg{A}\otimes\Alg{B}$ given by
\begin{equation}
\label{eq:canonicalSOT}
\map{E}\star \rho=\frac{1}{2}\big\{\rho\otimes \mathds{1}_{B}\,,\Jamiol[\mathcal{E}]\big\}\,, 
\end{equation}
where $\{*\,,*\}$ denotes the anti-commutator. 
\ed

In accordance with Refs.~\cite{HHPBS17,FuPa22,FuPa22a,LiNg23,PFBC23,FuPa24a}, we refer to $\map{E}\star \rho$ as a \emph{quantum state over time} associated with the process $(\map{E},\rho)$. While alternative formulations of a $\star$-product $\map{E}\star \rho$ have appeared in the literature \cite{LeSp13,HHPBS17}, the spatiotemporal product as given by \eqref{eq:canonicalSOT} is a spatiotemporal analogue of a bipartite density operator that has been characterized from different perspectives in Refs.~\cite{LiNg23,PFBC23,FuPa24a}, and whose physical significance has been addressed in Refs.~\cite{BDOV13,BDOV14,HHPBS17,FuPa22,FuPa24a}. Furthermore, the spatiotemporal product is 
used for the second notion of inverse needed for our main theorem~\cite{FuPa22a}. 

\bd
\label{defn:bayesmap}
Let $\map{E}:\Alg{A}\to \Alg{B}$ be a quantum channel and let $\rho\in\Alg{A}$ be a state. A quantum channel $\map{F}:\Alg{B}\to \Alg{A}$ is said to be a \define{Bayesian inverse} of the process $(\map{E},\rho)$ iff 
\begin{align}\label{BXSRLX67}
\map{E}\star \rho=\mathcal{S}\big(\map{F} \star \map{E}(\rho)\big) \, ,
\end{align}
where $\mathcal{S}:\Alg{B}\otimes \Alg{A}\to \Alg{A}\otimes \Alg{B}$ is the \define{swap map} given by the linear extension of the assignment $\mathcal{S}(\sigma\otimes \tau)=\tau\otimes \sigma$. 
\ed

The above definition of Bayesian inverse is based on extending the notion of reversibility to open systems beyond unitary dynamics~\cite{Ts22,FuPa22a,SASDS23}, and more generally beyond unital (i.e., bistochastic) quantum channels~\cite{CGS17,CAZ21,ChLi22}. Not every quantum process admits a Bayesian inverse, though many important systems do, including unitary channels, thermal operations~\cite{AWWW18}, unital quantum channels, Davies maps~\cite{Davies74,RFZ10}, perfect quantum error-correcting codes~\cite{PaBayes}, and more.  

\section{Results}

\subsection{Main Theorem}

\bt\label{MTX87}
Let $A$ and $B$ be quantum systems, and let $\mathscr{S}=\mathscr{L}_A\times \mathscr{L}_B$. Then the notions of Bayesian inverse and $\mathscr{S}$-operational inverse are equivalent.
\et

To expound on Theorem~\ref{MTX87}, recall that an element of $\mathscr{S}$ is a pair $(\obs{A},\obs{B})$ with $\obs{A}$ and $\obs{B}$ light-touch observables. 
The \emph{equivalence} in Theorem~\ref{MTX87} means that $\map{F}:\Alg{B}\to\Alg{A}$ is a Bayesian inverse of $(\map{E},\rho)$ if and only if $\map{F}$ is an $\mathscr{S}$-operational inverse inverse of $(\map{E},\rho)$. The utility of Theorem~\ref{MTX87} lies in the fact that given a quantum process $(\map{E},\rho)$, one may use equation~\eqref{BXSRLX67} in the definition of Bayesian inverse to \emph{solve} for an $\mathscr{S}$-operational inverse of ($\map{E},\rho)$. 
Such a solution $\map{F}$ then necessarily yields two-time expectation values such that for all pairs of light-touch observables $\obs{A}$ and $\obs{B}$, 
\be \label{TSXESX247}
\langle \obs{A}\, , \obs{B} \rangle=\langle \obs{B}\, , \obs{A} \rangle\, ,
\ee
where $\langle \obs{A}\, , \obs{B} \rangle$ and $\langle \obs{B}\, , \obs{A} \rangle$ are the two-time expectation values associated with the processes $(\map{E},\rho)$ and $(\map{F},\map{E}(\rho))$, respectively. On the other hand, trying to determine an $\mathscr{S}$-operational inverse from the definition alone would be impractical, as it would require finding a channel $\map{F}$ such that \eqref{TSXESX247} holds \emph{for all} pairs of light-touch observables $(\obs{A},\obs{B})\in \mathscr{S}$.

As a special case of Theorem~\ref{MTX87}, we first address the case of unitary evolution associated with a closed quantum system~\cite{Fritz10}. 
In particular, if $\Alg{A}=\Alg{B}$ and $\map{E}=\Ad_{U}$ with $U\in\Alg{A}$ unitary (so that $\map{E}(\sigma)=U\sigma U^{\dag}$ for all $\sigma\in\Alg{A}$), Ref.~\cite{FuPa22a} showed that the Bayesian inverse of $(\map{E},\rho)$ is $\map{E}^{-1}=\Ad_{U^{\dag}}$ for any state $\rho\in\Alg{A}$, as one would expect. 
To show $\map{E}^{-1}$ is also an $\mathscr{S}$-operational inverse, we first consider the case of dichotomic observables $\obs{A}$ and $\obs{B}$, so that $\obs{A}=P-(\mathds{1}_{A}-P)=2P-\mathds{1}_{A}$ and $\obs{B}=Q-(\mathds{1}_{B}-Q)=2Q-\mathds{1}_{B}$, with $P$ and $Q$ orthogonal projections onto the $+1$ eigenspaces of $\obs{A}$ and $\obs{B}$, respectively. Expanding out the two-time expectation values using~\eqref{2TEXT57} yields
\begin{align}
\<\obs{A}\, &,\obs{B}\>=\Tr\Big[\map{E}\big(P\rho P-P^{\perp}\rho P^{\perp}\big)(2Q-\mathds{1}_{B})\Big] \nonumber\\
&=\Tr\Big[\big(\{P,\rho\}-\rho\big)\big(2\map{E}^{-1}(Q)-\mathds{1}_{A}\big)\Big] \nonumber\\
&=\Tr\Big[2\{P,\rho\}\map{E}^{-1}(Q)-2\rho\map{E}^{-1}(Q)-\{P,\rho\}+\rho\Big] \nonumber\\
&=\Tr\Big[2\big\{\map{E}^{-1}(Q),\rho\big\}P-2\rho P-\big\{\map{E}^{-1}(Q),\rho\big\}+\rho\Big] \nonumber\\
&=\Tr\Big[\big(\big\{\map{E}^{-1}(Q),\rho\big\}-\rho\big)(2P-\mathds{1}_{A})\Big]\nonumber \\
&=\Tr\Big[\map{E}^{-1}\big(Q\map{E}(\rho)Q-Q^{\perp}\map{E}(\rho)Q^{\perp}\big)(2P-\mathds{1}_{A})\Big]\nonumber \\
&=\<\obs{B}\, ,\obs{A}\>, \label{eq:proofforunitaries}
\end{align}
where $P^{\perp}=\mathds{1}_{A}-P$ and $Q^{\perp}=\mathds{1}_{B}-Q$. In~\eqref{eq:proofforunitaries}, we used the cyclicity of trace as well as properties of $\map{E}$ and its reverse process $\map{E}^{-1}=\Ad_{U^{\dag}}$. 
For general light-touch observables that are not scalar multiples of the identity matrix, the result then follows from the fact that $\langle \lambda \obs{A}\,, \mu \obs{B}\rangle=\lambda \mu\langle \obs{A}\,,  \obs{B}\rangle$. The case when either $\obs{A}$ or $\obs{B}$ is a scalar multiple of the identity matrix also follows by direct calculation. As the proof of Theorem~\ref{MTX87} for general open quantum systems utilizes some technical results proved in Ref.~\cite{FuPa24a}, it is left to Appendix~\ref{sec:proofofmain}.

In the next section, we put Theorem~\ref{MTX87} to use by explicitly  constructing $\mathscr{S}$-operational inverses for open quantum systems that may be modeled by an amplitude-damping channel, which we show indeed yields time-symmetric expectation values for light-touch observables.

\subsection{Example: Amplitude-damping channel}
\label{sec:exADC}

\begin{figure}
\includegraphics[width=4.0cm]{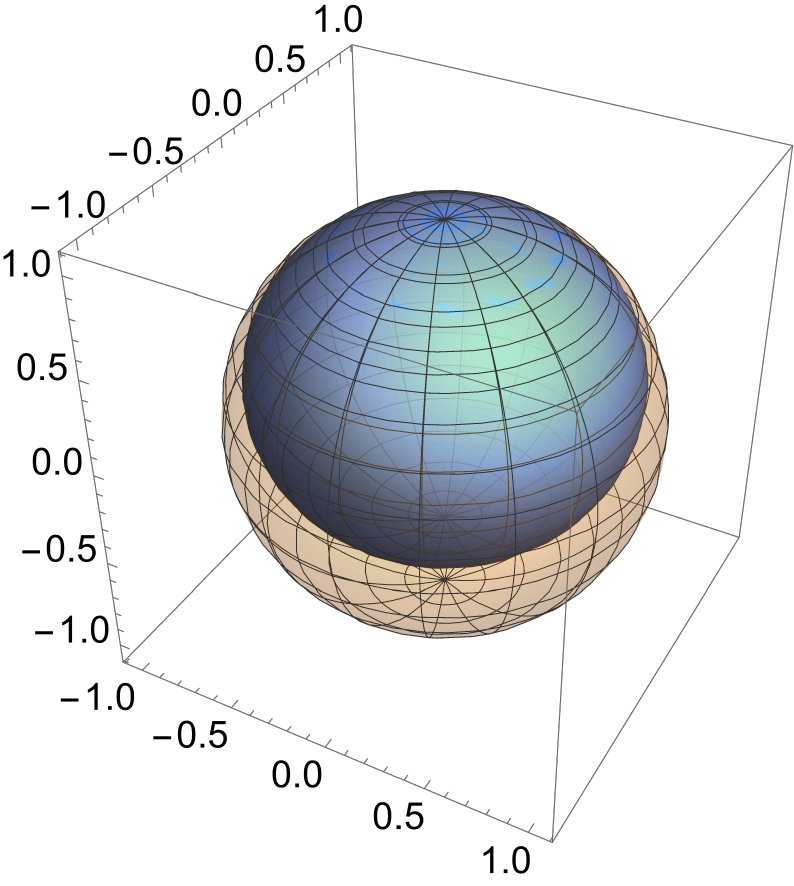}
\;\;\;
\includegraphics[width=4.0cm]{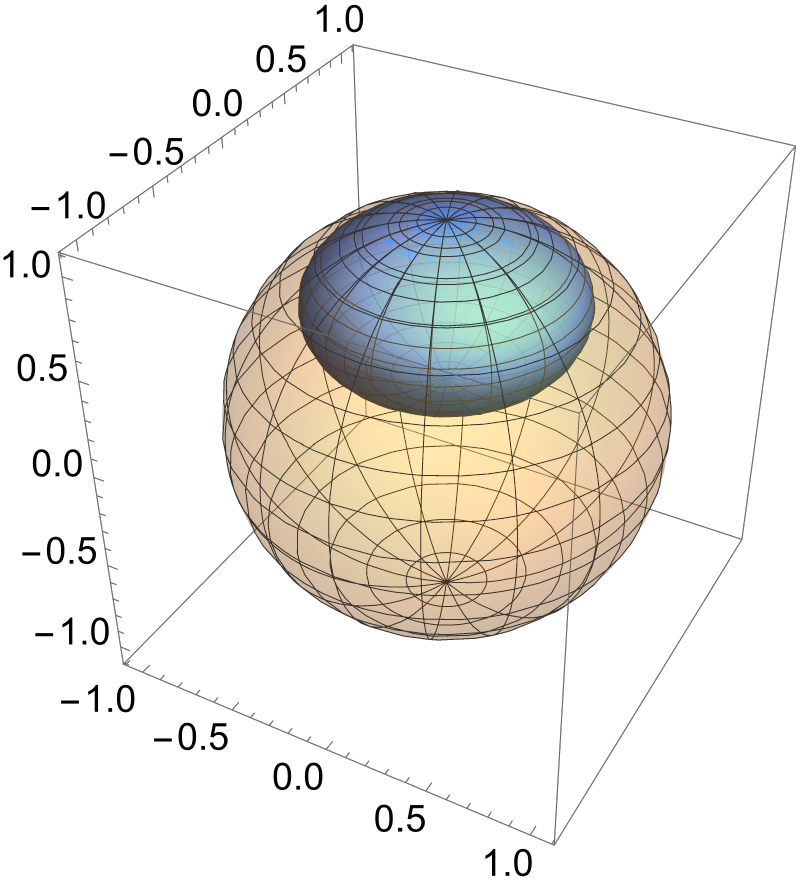}
\caption{Visualizations of the image (in blue) of the Bloch ball (in translucent orange) after one application of the amplitude-damping channel $\map{E}$ with $\gamma=0.2$ and $\gamma=0.6$, respectively. 
The north pole on the Bloch ball corresponds to the pure state $|0\>\<0|$, which is fixed under $\map{E}$.}
\label{fig:ADCvisual}
\end{figure}

The \emph{amplitude-damping} (or \emph{spontaneous decay}) \emph{channel} with \emph{damping parameter} $\gamma\in[0,1]$ is given by
\begin{equation}
\map{E}=\Ad_{E_0}+\Ad_{E_1},
\end{equation}
where 
\begin{equation}
\label{eq:ADCKrausOperators}
E_{0}=\begin{pmatrix}1&0\\0&\sqrt{1-\gamma}\end{pmatrix}
\quad\text{ and }\quad
E_{1}=\begin{pmatrix}0&\sqrt{\gamma}\\0&0\end{pmatrix}
\end{equation}
(recall that $\Ad_{E_i}(a)=E_i aE_i^{\dag}$ for all $a\in \Alg{A}$)~\cite{NiCh11}.
A visualization of what happens to the Bloch ball after one application of the amplitude-damping channel is shown in Figure~\ref{fig:ADCvisual} for the cases $\gamma=0.2$ and $\gamma=0.6$. 

Given a process $(\map{E},\rho)$, it was shown in Ref.~\cite{FuPa22a} that a Bayesian inverse $\map{F}$ of $(\map{E},\rho)$ necessarily satisfies
\be\label{eq:Binvgeneralformula}
\map{F}\big(|w_{k}\>\<w_{l}|\big)=\frac{\big\{\rho,\map{E}^*\big(|w_{k}\>\<w_{l}|\big)\big\}}{q_{k}+q_{l}}
\ee
for all $k,l$,
where 
\begin{equation}
\map{E}(\rho)=\sum_{k}q_{k}|w_{k}\>\<w_{k}|
\end{equation}
is an eigendecomposition of $\map{E}(\rho)$ using rank-1 projectors and $\map{E}^*:\Alg{B}\to\Alg{A}$ is the \emph{Hilbert--Schmidt adjoint} of $\map{E}$, which is defined to be the unique linear map satisfying
\be
\Tr\big[\map{E}(a)^{\dag}b\big]=\Tr\big[a^{\dag}\map{E}^*(b)\big]
\ee
for all $a\in\Alg{A}$ and $b\in\Alg{B}$.

We first consider a diagonal density matrix of the form 
\begin{equation}
\label{eq:priorrhoADC}
\rho=\frac{1}{2}\begin{pmatrix}1+r_3&0\\0&1-r_3\end{pmatrix}
\end{equation}
so that 
\begin{equation}
\label{eq:predictionrhoADC}
\map{E}(\rho)=\frac{1}{2}\begin{pmatrix}1+s_3&0\\0&1-s_3\end{pmatrix},
\end{equation}
where 
\begin{equation}
\label{eq:s3}
s_{3}=r_3+\gamma(1-r_3)\, .
\end{equation}
This expresses the density matrices in the computational basis $|0\>,|1\>$. 
From this, we can read off 
\begin{equation}
q_{0}=
\frac{1+s_3}{2}
\quad\text{ and }\quad
q_{1}=
\frac{1-s_3}{2}\, ,
\end{equation}
so that 
\begin{equation}
\map{E}(\rho)=q_{0}|0\>\<0|+q_{1}|1\>\<1|\, .
\end{equation}

\begin{figure}
\includegraphics[width=4.0cm]{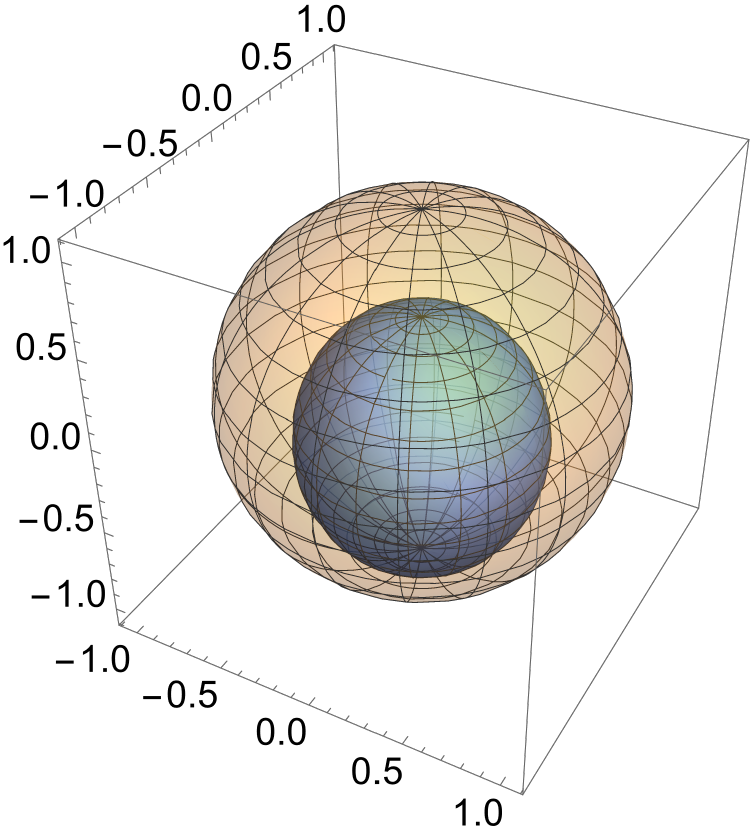}
\;\;\;
\includegraphics[width=4.0cm]{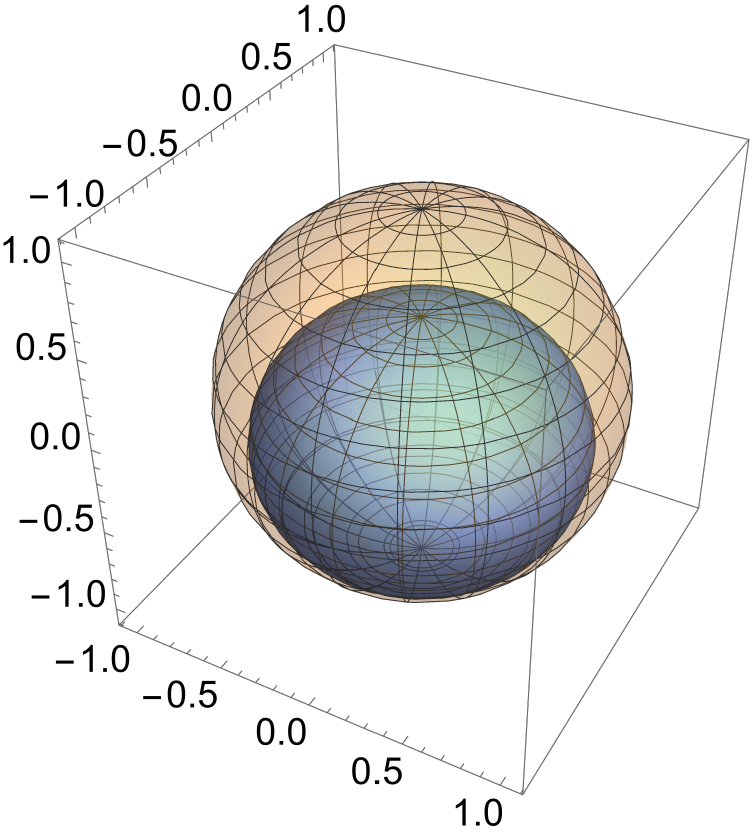}
\caption{A visualization of the image (in blue) of the Bloch ball (in translucent orange) after one application of the Bayesian inverse $\map{F}$ of the amplitude-damping channel with $r_3=0.2$ and $\gamma=0.6$ is shown on the left. Meanwhile, the right shows the analogous situation but with the Petz recovery map $\map{R}$ for the same values of $r_{3}$ and $\gamma$.
While both channels are bit-flipped amplitude-damping channels, the Bayesian inverse has an additional dephasing effect.}
\label{fig:ADCBayesVSPetz}
\end{figure}

The map $\map{F}$ as defined by~\eqref{eq:Binvgeneralformula} is completely positive whenever (cf.\ Theorem~\ref{thm:ADCsymmetry} in Appendix~\ref{app:ADC})
\be
\label{eq:r3gammaconstraint}
r_{3}\ge\frac{\gamma}{\gamma-2}\, ,
\ee
which is automatically satisfied if $r_{3}\ge0$. Under condition~\eqref{eq:r3gammaconstraint}, the Bayesian inverse is given by 
\begin{equation}
\label{eq:BayesianinverseADC}
\map{F}=\Ad_{F_{0}}+\Ad_{F_{1}}+\Ad_{F_{2}},
\end{equation}
where the Kraus operators are
\begin{equation}
F_{0}=
\begin{pmatrix}
\sqrt{\frac{1+r_3}{1+s_3}}&0\\
0&\sqrt{\frac{(1-\gamma)(1+s_3)}{1+r_3}}
\end{pmatrix}
,
\end{equation}
\begin{equation}
F_{1}=
\begin{pmatrix}
0&0\\
\sqrt{\frac{\gamma(1-r_3)}{1+s_3}}&0
\end{pmatrix}
\!, \text{ and }
F_{2}=
\begin{pmatrix}0 & 0 \\
0&\sqrt{\frac{\gamma(r_3+s_3)}{1+r_3}}
\end{pmatrix}\!.
\end{equation}
As discussed in Appendix~\ref{app:ADC}, the Bayesian inverse is a 
bit-flipped amplitude-damping channel combined with a dephasing channel (see also the left graphic in Figure~\ref{fig:ADCBayesVSPetz}). 

In Tables~\ref{tab:ADCTTEVFPB} and~\ref{tab:ADCBayesTTEVFPB}, we list the two-time expectation values of Pauli observables associated with the processes $(\map{E},\rho)$ and $(\map{F},\map{E}(\rho))$, respectively. As the two tables differ by a transposition, such two-time expectation values are indeed time-symmetric, in accordance with Theorem~\ref{MTX87}.

\begin{table}
\begin{tabular}{c|cccc}
&$\sigma_{0}$&$\sigma_{1}$&$\sigma_{2}$&$\sigma_{3}$\\
\hline
$\sigma_{0}$ & $1$ & $0$ & $0$ & $r_3+\gamma(1-r_3)$ \\
$\sigma_{1}$ & $0$ & $\sqrt{1-\gamma}$ & $0$ & $0$ \\
$\sigma_{2}$ & $0$ & $0$ & $\sqrt{1-\gamma}$ & $0$ \\
$\sigma_{3}$ & $r_3$ & $0$ & $0$ & $1-\gamma(1-r_3)$\\
\end{tabular}
\caption{The two-time expectation values for the amplitude-damping channel $\map{E}$ with initial density matrix $\rho$ from~\eqref{eq:priorrhoADC}. Each row corresponds to a measurement for Alice (the first measurement performed) and each column corresponds to a measurement for Bob (the second measurement performed, which occurs after the evolution $\map{E}$). In this way, the table specifies a matrix whose $(\alpha,\beta)$ entry is $\<\sigma_{\alpha},\sigma_{\beta}\>$, which can be efficiently computed using~\cite[Theorem 5.3]{FuPa24a} (see also Lemma~\ref{MTXS45739} in Appendix~\ref{sec:proofofmain}).}
\label{tab:ADCTTEVFPB}
\end{table}

\begin{table}
\begin{tabular}{c|cccc}
&$\sigma_{0}$&$\sigma_{1}$&$\sigma_{2}$&$\sigma_{3}$\\
\hline
$\sigma_{0}$ & $1$ & $0$ & $0$ & $r_3$ \\
$\sigma_{1}$ & $0$ & $\sqrt{1-\gamma}$ & $0$ & $0$ \\
$\sigma_{2}$ & $0$ & $0$ & $\sqrt{1-\gamma}$ & $0$ \\
$\sigma_{3}$ & $r_3+\gamma(1-r_3)$ & $0$ & $0$ & $1-\gamma(1-r_3)$\\
\end{tabular}
\caption{The two-time expectation values for the Bayesian inverse $\map{F}$ of the amplitude-damping channel $\map{E}$ with initial density matrix $\rho$ from~\eqref{eq:priorrhoADC}. Each row corresponds to a fixed measurement for Bob (first measurement) and each column corresponds to a fixed measurement for Alice (second measurement). In this way, the table specifies a matrix whose $(\beta,\alpha)$ entry is $\<\sigma_{\beta},\sigma_{\alpha}\>$. By Theorem~\ref{MTX87}, this matrix is the transpose of the matrix from Table~\ref{tab:ADCTTEVFPB}.}
\label{tab:ADCBayesTTEVFPB}
\end{table}

It is worth comparing the Bayesian inverse $\map{F}$ with the Petz recovery map $\map{R}:\Alg{B}\to\Alg{A}$ of the process $(\map{E},\rho)$~\cite{Connes74,Pe84,Petz86,Cr08,Le06,Le07,BaKn02}, 
which is given by 
\be
\map{R}(\sigma)=\rho^{1/2}\map{E}^*\big(\map{E}(\rho)^{-1/2}\sigma\map{E}(\rho)^{-1/2}\big)\rho^{1/2}. 
\ee
The Petz recovery map, which is often used for time-reversal symmetry, retrodiction, relative entropy inequalities, thermodynamics, and approximate error-correction~\cite{BaKn02,Pe03,NgMa10,MaNg12,Wilde15,JRSWW16,BDW16,SuToHa16,LiWi18,AWWW18,JRSWW18,CHPSSW19,CPS20,BuSc21,AwBuSc21,KMK22,LdMB22,PaBu22,BaVa23,Pa24,LJPB24,AZBS24,Ts24,LuYu24}, is completely positive for all $r_{3}\in(-1,1)$ and $\gamma\in(0,1)$ and in this example corresponds to a bit-flipped amplitude-damping channel (with no dephasing). 
A comparison of the action of the Bayesian inverse $\map{F}$ versus the Petz recovery map $\map{R}$ on the Bloch ball is shown in Figure~\ref{fig:ADCBayesVSPetz}. 
A more in-depth analysis of the Petz recovery map $\map{R}$ and how it differs from the Bayesian inverse $\map{F}$ is discussed in Appendix~\ref{app:ADC}. In short, the Petz recovery map \emph{does not} reproduce time-symmetric expectation values for the amplitude-damping channel. 

\subsection{Experimental proposal}

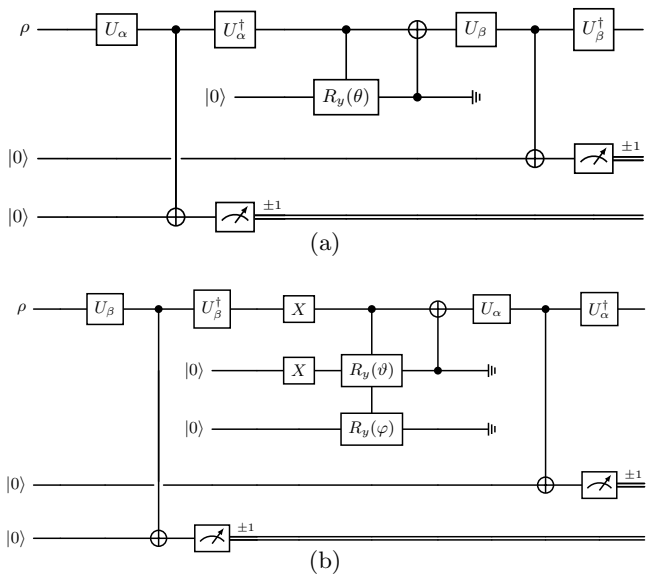
\begin{figure}
\centering
\begin{tikzpicture}
\node[scale=0.78] at (0,0) {
\begin{quantikz}
\setwiretype{n} & \lstick{$\rho$} & \setwiretype{q} & \gate{U_{\alpha}} & \ctrl{3} & \gate{U_{\alpha}^{\dag}} & & \ctrl{1} & \targ{} & \gate{U_{\beta}} & \ctrl{2} & \gate{U_{\beta}^{\dag}} & \\ 
\setwiretype{n}  & & & & & \lstick{\ket{0}} & \setwiretype{q} & \gate{R_{y}(\theta)} & \ctrl{-1} & \ground{} \\
\setwiretype{n} & \lstick{\ket{0}} & \setwiretype{q} & & \push{\phantom{x}} & & & & & & \targ{} & \meter{} & \setwiretype{c} \wire[l][1]["\pm 1"{above,pos=0.4}]{a} \\
\setwiretype{n} & \lstick{\ket{0}} & \setwiretype{q} & & \targ{}\wire[u][3]{q} & \meter{} & \setwiretype{c} \wire[l][1]["\pm 1"{above,pos=0.4}]{a} & & & & & &
\end{quantikz}
};
\node at (0,-1.65) {(a)};
\node[scale=0.71] at (0,-4) {
\begin{quantikz}
\setwiretype{n} & \lstick{$\rho$} & \setwiretype{q} & \gate{U_{\beta}} & \ctrl{3} & \gate{U_{\beta}^{\dag}} & & \gate{X} & \ctrl{2} & \targ{} & \gate{U_{\alpha}} & \ctrl{3} & \gate{U_{\alpha}^{\dag}} & \\ 
\setwiretype{n}  & & & & & \lstick{\ket{0}} & \setwiretype{q} & \gate{X} & \gate{R_{y}(\vartheta)} & \ctrl{-1} & \ground{} \\
\setwiretype{n}  & & & & & \lstick{\ket{0}} & \setwiretype{q} & & \gate{R_{y}(\varphi)} & & \ground{}  \\
\setwiretype{n} & \lstick{\ket{0}} & \setwiretype{q} & & \push{\phantom{x}} & & & & & & & \targ{} & \meter{} & \setwiretype{c} \wire[l][1]["\pm 1"{above,pos=0.4}]{a} \\
\setwiretype{n} & \lstick{\ket{0}} & \setwiretype{q} & & \targ{}\wire[u][3]{q} & \meter{} & \setwiretype{c} \wire[l][1]["\pm 1"{above,pos=0.4}]{a} & & & & & & &
\end{quantikz}
};
\node at (0,-5.875) {(b)};
\end{tikzpicture}
\caption{Quantum circuits are read from left to right in time~\cite{qiskit2024}. 
The unitary operators $U_{\gamma}$ associated with the measurement of the Pauli operators $\sigma_{\gamma}$ are given by $U_{1}=H$, $U_{2}=HR_{z}(\pi/2)$, and $U_{3}=\mathds{1}_{2}$, where $H=\frac{1}{\sqrt{2}}\left(\begin{smallmatrix}1&1\\1&-1\end{smallmatrix}\right)$ is the Hadamard gate, $R_{y}(\theta)=\left(\begin{smallmatrix}\cos(\theta/2)&-\sin(\theta/2)\\\sin(\theta/2)&\cos(\theta/2)\end{smallmatrix}\right)$ is the $SU(2)$ rotation operator about the $y$-axis for $\theta\in[0,4\pi)$, and $R_{z}(\delta)=\left(\begin{smallmatrix}1&0\\0&e^{i\delta}\end{smallmatrix}\right)$ is the phase-shift gate for $\delta\in [0,2\pi)$~\cite{BCS04}. 
Since $\sigma_{0}=\mathds{1}_{2}$ corresponds to no measurement, this case is not illustrated. 
(a) The circuit depicts the action of a sequential measurement of $\sigma_{\alpha}$ followed by $\sigma_{\beta}$ with an amplitude-damping channel $\map{E}$ between measurements.
The parameter $\theta$ is related to the amplitude-damping parameter $\gamma$ by $\cos\left(\frac{\theta}{2}\right)=\sqrt{1-\gamma}$.
(b) The circuit depicts the action of a sequential measurement of $\sigma_{\beta}$ followed by $\sigma_{\alpha}$ with the Bayesian inverse $\map{F}$ of $(\map{E},\rho)$ between measurements.
The parameters $\vartheta$ and $\varphi$ are related to $r_{3}$ and $\gamma$ by $\cos\left(\frac{\vartheta}{2}\right)=\sqrt{\frac{1+r_{3}}{1+s_{3}}}$ and $\cos\left(\frac{\varphi}{2}\right)=\sqrt{\frac{(1-\gamma)(1+s_3)}{1+r_3}}$, where $s_{3}$ is as in~\eqref{eq:s3}.
}
\label{fig:twotimemeasurementcircuit}
\end{figure}

In Figure~\ref{fig:twotimemeasurementcircuit} (a), we depict a quantum circuit that performs a measurement of a Pauli observable $\sigma_{\alpha}$ on the input state $\rho=\frac{1}{2}(\mathds{1}_{2}+r_{3}\sigma_{3})$, exposes the updated qubit to amplitude-damping noise $\map{E}$, and then measures another Pauli observable $\sigma_{\beta}$. 
The channel $\map{E}$ is represented by a Stinespring ampliation~\cite{St55,Pa18} coupling the system to an ancillary qubit followed by unitary evolution and then discarding the environment output state~\cite{NiCh11,TDNFC23}.
Figure~\ref{fig:twotimemeasurementcircuit} (b) depicts the proposed operational time-reversed circuit involving the Bayesian inverse $\map{F}$ of $(\map{E},\rho)$.
The channel $\map{F}$ is a bit-flipped amplitude-damping and dephasing channel represented as a circuit involving two ancillary qubits.
The operational time-reversed circuit measures $\sigma_{\beta}$ on the input state $\map{E}(\rho)=\frac{1}{2}(\mathds{1}_{2}+s_{3}\sigma_{3})$, exposes the updated qubit to the noise $\map{F}$, and measures $\sigma_{\alpha}$.
The circuits in Figure~\ref{fig:twotimemeasurementcircuit} permit the usage of the same qubit in multiple stages of the protocol~\cite{ICKHC16,ICC17,LlVi01,AnOi08,SNAKDSGJ17}. 
Based on recent experimental simulations of the amplitude-damping, dephasing, and depolarizing channels using nuclear magnetic resonance~\cite{XWPSL17} and photon path states~\cite{MMPGESP15}, it seems plausible that our proposed experiments can be tested on current technologies.

The example of the amplitude-damping channel provides a method to test the operational time-reversal symmetry of two-time expectation values for a quantum channel that is neither unitary nor bistochastic (i.e., a unital quantum channel). Namely, for a qubit density matrix of the form $\rho=\frac{1}{2}(\mathds{1}_{2}+r_{3}\sigma_{3})$ with $r_{3}\in(-1,1)$ and an amplitude-damping channel $\map{E}$ determined by amplitude-damping parameter $\gamma\in(0,1)$, such time-reversal symmetry is possible when $r_{3}>\frac{\gamma}{\gamma-2}$. 
However, $(\map{E},\rho)$ is no longer Bayesian invertible if the initial state $\rho$ is perturbed from a diagonal density matrix to one that is not diagonal. This means that if there is any error in the state-preparation process, then it would seem that experimentally testing our predictions of two-time expectation value time-reversal symmetry is not possible. Fortunately, by adding an infinitesimal amount of noise to the amplitude-damping channel, say, in the form of completely depolarizing noise, one can find an open set of initial states $\rho$ for which $(\map{E},\rho)$ is indeed Bayesian invertible. 
This illustrates the robustness of Bayesian inverses under perturbations so that one could, in principle, experimentally test the predictions set forth in this work (see Theorem~\ref{thm:ADCrobustness} and Appendix~\ref{app:ADC} for details).

\section{Discussion and Conclusion}

In this work, we introduced the notion of an $\mathscr{S}$-operational inverse of a quantum channel $\map{E}:\Alg{A}\to \Alg{B}$ with respect to an input state $\rho$. It is defined to be a channel $\map{F}:\Alg{B}\to \Alg{A}$ that realizes time-symmetric expectation values for all pairs of observables $(\obs{A},\obs{B})\in \mathscr{S}$. When $\mathscr{S}$ consists of all pairs of light-touch observables, we showed that $\mathscr{S}$-operational inverses coincide with Bayesian inverses as introduced in Ref.~\cite{FuPa22a}. Due to the fact that Bayesian inverses are often computable, such a result allows one to explicitly determine $\mathscr{S}$-operational inverses associated with general quantum channels. Making use of our results, we then explicitly constructed $\mathscr{S}$-operational inverses for amplitude-damping channels, and we showed how such maps indeed realize time-symmetric expectation values for suitable choices of input states.

Our results suggest an operational time-reversal symmetry for quantum systems beyond the setting of unitary (and even bistochastic) evolution. Such a temporal symmetry for open quantum systems, which may leak information to their environment, is quite striking, and has yet to be explored in more detail. For example, given an arbitrary initial state $\rho$ and quantum channel $\map{E}$, under what necessary and sufficient conditions does a Bayesian inverse exist? Moreover, the operational nature of our results naturally lend themselves to testable predictions by experiment, and thus provide a fertile testing ground for the quantum Bayes' rule proposed in Ref.~\cite{FuPa22a}.

\vspace{3mm}
\noindent
\textbf{Acknowledgements}. AJP thanks Rob Spekkens, Jacopo Surace, and Y\`il\`e Y\={\i}ng for discussions and the Perimeter Institute for their hospitality during a visit. 
This research was supported in part by Perimeter Institute for Theoretical Physics. Research at Perimeter Institute is supported by the Government of Canada through the Department of Innovation, Science and Economic Development and by the Province of Ontario through the Ministry of Research, Innovation and Science. 
All figures besides Figure~\ref{fig:twotimemeasurementcircuit} were created in Mathematica~\cite{Mathematica2024}.

\bibliography{references}

\clearpage
\newpage

\title{Methods}
\author{testing}

\maketitle
\onecolumngrid
\vspace{1cm}

\begin{center}\large \textbf{Time-symmetric correlations for open quantum systems} \\
\textbf{--- Supplementary Material ---}\\
\end{center}


\vspace{-5mm}

 \appendix

\section{Classical time-reversal symmetry of sequential measurements}
\label{sec:classicalBayes}

Here, we recall the time symmetry associated with the measurements of classical random variables in order to provide a clearer comparison with our definition of operational inverse. 
Let $X$ and $Y$ be two random variables on a finite population, and let $\mathbb{P}(x):=\mathbb{P}(X=x)$ and $\mathbb{P}(y):=\mathbb{P}(Y=y)$ denote the probabilities of measuring the values $x$ and $y$, respectively. The joint probability of measuring $x$ and $y$ will be denoted by $\mathbb{P}(x,y)$. This joint probability can be equivalently expressed in two different ways. 
On the one hand, it can be expressed as the joint probability of first measuring $X$ and then $Y$ via the expression 
\be
\mathbb{P}(x,y)=\mathbb{P}(y|x)\mathbb{P}(x)\, ,
\ee
where $\mathbb{P}(y|x)$ denotes the conditional probability of outcome $y$ being measured given that $x$ was the outcome of measuring $X$. 
Alternatively, it can be expressed as the joint probability of first measuring $Y$ and then $X$ via the expression 
\be
\mathbb{P}(x,y)=\mathbb{P}(x|y)\mathbb{P}(y)\, ,
\ee
where $\mathbb{P}(x|y)$ denotes the conditional probability of outcome $x$ being measured given that $y$ was the outcome of measuring $Y$. 
It then follows that the expected value of the \emph{product} of the measurements $X$ followed by $Y$ is given by 
\be
\<X,Y\>=\sum_{y}\sum_{x}xy\mathbb{P}(y|x)\mathbb{P}(x)\, ,
\ee
and similarly the expected value of the product of the measurements $Y$ followed by $X$ is given by  
\be
\<Y,X\>=\sum_{x}\sum_{y}yx\mathbb{P}(x|y)\mathbb{P}(y)\, .
\ee
By Bayes' rule, 
\be
\mathbb{P}(y|x)\mathbb{P}(x)=
\mathbb{P}(x|y)\mathbb{P}(y)\, ,
\ee
from which it follows that $\<X,Y\>=\<Y,X\>$. 

\section{Proof of Theorem~\ref{MTX87}}
\label{sec:proofofmain}
In this section we prove Theorem~\ref{MTX87}, which uses two results from Ref.~\cite{FuPa24a} that we state as lemmas. 

\blem[\cite{FuPa24a} Theorem 5.3]
\label{MTXS45739}
Let $\map{E}:\Alg{A}\to \Alg{B}$ be a quantum channel with initial state $\rho\in \Alg{A}$. Then, for all light-touch observables $\obs{A}$ on $A$ and $\obs{B}$ on $B$, 
\begin{align}
\langle \obs{A} \, , \obs{B}\rangle=\Tr\Big[(\map{E}\star \rho)(\obs{A}\otimes \obs{B})\Big]\, ,
\end{align}
where $\langle \obs{A} \, , \obs{B}\rangle$ is the two-time expectation value with respect to $(\map{E},\rho)$ and $\map{E}\star \rho$ is the spatiotemporal product of $\map{E}$ and $\rho$ (as defined by \eqref{eq:canonicalSOT}).
\elem

\blem[\cite{FuPa24a} Proposition~5.8]
\label{LMTXS35739}
Given a finite-dimensional quantum system $A$, there exists a basis of the real vector space $\mathbf{Obs}(A)$ consisting of light-touch observables.
\elem

We now give the proof of our main theorem.

\begin{proof}[Proof of Theorem~\ref{MTX87}]
Let $\map{E}:\Alg{A}\to \Alg{B}$ be a quantum channel with initial state $\rho\in \Alg{A}$, let $\mathscr{S}=\mathscr{L}_A\times \mathscr{L}_B\subset \mathbf{Obs}(A)\times \mathbf{Obs}(B)$, let $\mathcal{S}:\Alg{B}\otimes \Alg{A}\to \Alg{A}\otimes \Alg{B}$ be the swap map, and suppose $\map{F}:\Alg{B}\to \Alg{A}$ is a Bayesian inverse of the process $(\map{E},\rho)$. Then for all pairs of light-touch observables $(\obs{A},\obs{B})\in \mathscr{S}$, 
\begingroup
\allowdisplaybreaks
\begin{align}
\langle \obs{A} \, , \obs{B}\rangle&=\Tr\Big[(\map{E}\star \rho)(\obs{A}\otimes \obs{B})\Big] && \text{by Lemma~\ref{MTXS45739}} \nonumber \\
&=\Tr\Big[\mathcal{S}\big(\map{F}\star \map{E}(\rho)\big)(\obs{A}\otimes \obs{B})\Big] && \text{by the definition of Bayesian inverse~\eqref{BXSRLX67}} \nonumber \\
&=\Tr\Big[\mathcal{S}\Big(\big(\map{F}\star \map{E}(\rho)\big)(\obs{B}\otimes \obs{A})\Big)\Big] && \text{since $\mathcal{S}$ is multiplicative} \nonumber \\
&=\Tr\Big[\big(\map{F}\star \map{E}(\rho)\big)(\obs{B}\otimes \obs{A})\Big] && \text{since $\mathcal{S}$ preserves trace} \nonumber \\
&=\langle \obs{B} \, , \obs{A}\rangle\, , && \text{by Lemma~\ref{MTXS45739}}  
\end{align}
\endgroup
where $\langle \obs{A} \, , \obs{B}\rangle$ and $\langle \obs{B} \, , \obs{A}\rangle$ denote the two-time expectation values with respect to the processes $(\map{E},\rho)$ and $(\map{F},\map{E}(\rho))$, respectively. Thus, $\map{F}$ is an $\mathscr{S}$-operational inverse of the process $(\map{E},\rho)$.

Now suppose $\map{F}:\Alg{B}\to \Alg{A}$ is an $\mathscr{S}$-operational inverse of the process $(\map{E},\rho)$, so that for all pairs of light-touch observables $(\obs{A},\obs{B})\in \mathscr{S}$,
\be \label{SMXEXT67}
\langle \obs{A} \, , \obs{B}\rangle=\langle \obs{B} \, , \obs{A}\rangle\, ,
\ee
where $\langle \obs{A} \, , \obs{B}\rangle$ and $\langle \obs{B} \, , \obs{A}\rangle$ denote the two-time expectation values with respect to the processes $(\map{E},\rho)$ and $(\map{F},\map{E}(\rho))$, respectively. Then for all pairs of light-touch observables $(\obs{A},\obs{B})\in \mathscr{S}$,
\begin{align}
\Tr\Big[(\map{E}\star \rho)(\obs{A}\otimes \obs{B})\Big]&=\langle \obs{A}\,, \obs{B} \rangle && \text{by Lemma~\ref{MTXS45739}} \nonumber \\
&=\langle \obs{B}\,, \obs{A} \rangle && \text{by equation \eqref{SMXEXT67}} \nonumber \\
&=\Tr\Big[\big(\map{F}\star \map{E}(\rho)\big)(\obs{B}\otimes \obs{A})\Big] && \text{by Lemma~\ref{MTXS45739}} \nonumber \\
&=\Tr\Big[\mathcal{S}\Big(\big(\map{F}\star \map{E}(\rho)\big)(\obs{B}\otimes \obs{A})\Big)\Big] && \text{since $\mathcal{S}$ preserves trace} \nonumber \\
&=\Tr\Big[\mathcal{S}\big(\map{F}\star \map{E}(\rho)\big)(\obs{A}\otimes \obs{B})\Big]\, . && \text{since $\mathcal{S}$ is multiplicative} 
\end{align}
Now by Lemma~\ref{LMTXS35739}, it follows that there exists bases $\{\obs{A}^{(i)}\}$ and $\{\obs{B}^{(j)}\}$  of the real vector spaces $\mathbf{Obs}(A)$ and $\mathbf{Obs}(B)$ such that $\obs{A}^{(i)}$ and $\obs{B}^{(j)}$ are light-touch observables for all $i$ and $j$, from which it follows that $\{\obs{A}^{(i)}\otimes \obs{B}^{(j)}\}$ is a basis of $\mathbf{Obs}(AB)$, where $AB$ is the joint system with associated algebra $\Alg{A}\otimes\Alg{B}$. Moreover, by the calculation above, it follows that
\be
\Tr\Big[(\map{E}\star \rho)\big(\obs{A}^{(i)}\otimes \obs{B}^{(j)}\big)\Big]
=\Tr\Big[\mathcal{S}\big(\map{F}\star \map{E}(\rho)\big)\big(\obs{A}^{(i)}\otimes \obs{B}^{(j)}\big)\Big]
\ee
for all $i$ and $j$. And since $\big\{\obs{A}^{(i)}\otimes \obs{B}^{(j)}\big\}$ is a basis of $\mathbf{Obs}(AB)$, it follows that 
\be
\map{E}\star \rho=\mathcal{S}\big(\map{F}\star \map{E}(\rho)\big)\, .
\ee
Thus, $\map{F}$ is a Bayesian inverse for the process $(\map{E},\rho)$. 
\end{proof}

\section{Amplitude-damping noise analysis}
\label{app:ADC}

This appendix serves several purposes. For one, it is meant to supply the calculations necessary to derive the results of Section~\ref{sec:exADC} (see Theorem~\ref{thm:ADCsymmetry} below). This appendix also includes comparisons between the Bayesian inverse and the Petz recovery map.  
Finally, this appendix contains the analysis for the robustness of Bayesian inverses for the amplitude-damping channel by looking at perturbations from a diagonal input density matrix to allow imperfect state preparation in order to be more readily accessible to experiment (see Theorem~\ref{thm:ADCrobustness} below). 

Let $\map{E}=\Ad_{E_0}+\Ad_{E_1}$ be the amplitude-damping channel as in Section~\ref{sec:exADC}. 
The Jamio{\l}kowski matrix associated with $\map{E}$ is then given by 
\begin{equation}
\Jamiol[\map{E}]=\begin{pmatrix}1&0&0&0\\0&0&\sqrt{1-\gamma}&0\\0&\sqrt{1-\gamma}&\gamma&0\\0&0&0&1-\gamma\end{pmatrix}.
\end{equation}
In addition to the Jamio{\l}kowski matrix (as  defined by Eq.~\eqref{JMX57}), we will also make use of the Choi matrix~\cite{Ch75}, which we now define.

\begin{definition}
\label{defn:Choi}
Let $\map{E}:\Alg{A}\to \Alg{B}$ be a quantum channel. The \define{Choi matrix} associated with $\map{E}$ is the element $\mathscr{C}[\map{E}]\in \Alg{A}\otimes \Alg{B}$ given by
\be
\mathscr{C}[\map{E}]=\sum_{i,j}|i\rangle \langle j|\otimes \map{E}(|i\rangle \langle j|)\, .
\ee
\end{definition}
The main utility of  the Choi matrix comes from the fact that a linear map $\map{E}:\Alg{A}\to\Alg{B}$ is completely positive if and only if $\Choi[\map{E}]$ is positive~\cite{Ch75}. We note, however, that while the Jamio{\l}kowski and Choi matrices have been defined in terms of the orthonormal basis $\{| i\rangle\}_{i=1}^{d}$ of the Hilbert space $\mathcal{H}_A$, only the Jamio{\l}kowski matrix is a basis-independent construction, as the Choi matrix is in fact basis-\emph{dependent} (a manifestly basis-independent definition of the Jamio{\l}kowski matrix was provided in Refs.~\cite{FuPa22,FuPa22a}).

Now, given an arbitrary input density matrix \begin{equation}
\label{eq:generalinitialqubitstate}
\rho=\frac{1}{2}\left(\mathds{1}+\vec{r}\cdot\vec{\sigma}\right)
=\frac{1}{2}\begin{pmatrix}1+r_{3}&r_{1}-ir_{2}\\r_{1}+ir_{2}&1-r_{3}\end{pmatrix},
\end{equation}
with $\vec{r}=(r_1,r_2,r_3)\in\R^{3}$ such that $\lVert\vec{r}\rVert\le1$ (this last condition is necessary so that $\rho\ge0$ or equivalently $\det(\rho)\ge0$), one can explicitly compute $\map{E}(\rho)$ as 
\begin{equation}
\map{E}(\rho)=\frac{1}{2}\left(\mathds{1}+\vec{s}\cdot\vec{\sigma}\right)\, ,
\end{equation}
where
\begin{equation}
\label{eq:Erhoimages}
\vec{s}=\left(\sqrt{1-\gamma}\,r_{1},\sqrt{1-\gamma}\,r_{2},r_{3}+\gamma(1-r_{3})\right). 
\end{equation}

As in Section~\ref{sec:exADC}, we henceforth assume $r_{1}=r_{2}=0$. The Choi and Jamio{\l}kowski matrices of the Bayesian inverse $\map{F}$ using formula~\eqref{eq:Binvgeneralformula} are then given by  
\begin{equation}
\label{eq:ChoiBayesmapr3}
\Choi[\map{F}]
=\begin{pmatrix}
\frac{1+r_3}{1+s_3}&0&0&\sqrt{1-\gamma}\\
0&\frac{\gamma(1-r_3)}{1+s_3}&0&0\\
0&0&0&0\\
\sqrt{1-\gamma}&0&0&1
\end{pmatrix}
\quad\text{ and }\quad
\Jamiol[\map{F}]
=\begin{pmatrix}
\frac{1+r_3}{1+s_3}&0&0&0\\
0&\frac{\gamma(1-r_3)}{1+s_3}&\sqrt{1-\gamma}&0\\
0&\sqrt{1-\gamma}&0&0\\
0&0&0&1
\end{pmatrix},
\end{equation}
respectively. 
The Choi matrix $\Choi[\map{F}]$ is Hermitian and its eigenvalues are all non-negative if $r_{3}\ge0$. 
Indeed, the eigenvalues are given by $\frac{\gamma(1-r_3)}{1+s_3}$, $0$, and the eigenvalues of the $2\times 2$ matrix
\be
\label{eq:r3gammasubmatrix}
\begin{pmatrix}\frac{1+r_3}{1+s_3}&\sqrt{1-\gamma}\\\sqrt{1-\gamma}&1\end{pmatrix}.
\ee
Since $\frac{1+r_{3}}{1+s_{3}}\ge0$, this latter matrix is positive if and only if its determinant is positive, which leads to the general constraint $r_{3}\ge\frac{\gamma}{\gamma-2}$ as in~\eqref{eq:r3gammaconstraint}, which holds for some negative values of $r_{3}$. 

As such, $\map{F}$ is completely positive and trace-preserving (CPTP) for all values of $\gamma\in(0,1),r_{3}\in[0,1)$.
In fact, by using the $LDL^{\dag}$ decomposition~\cite{Strang2022}, one can factorize the Choi matrix as $\Choi[\map{F}]=XX^{\dag}$, where $X=LD^{1/2}$ is 
\begin{equation}
\label{eq:XforBayesmapr3}
X=
\left(\begin{array}{cccc}
\cellcolor{red!10}\sqrt{\frac{1+r_3}{1+s_3}}&\cellcolor{yellow!10}0&0&\cellcolor{blue!10}0\\
\cellcolor{red!10} 0&\cellcolor{yellow!10}\sqrt{\frac{\gamma(1-r_3)}{1+s_3}}&0&\cellcolor{blue!10}0\\
\cellcolor{orange!10}0&\cellcolor{green!10}0&0&\cellcolor{violet!10}0\\
\cellcolor{orange!10}\sqrt{\frac{(1-\gamma)(1+s_3)}{1+r_3}}&\cellcolor{green!10}0&0&\cellcolor{violet!10}\sqrt{\frac{\gamma(r_3+s_3)}{1+r_3}}
\end{array}\right).
\end{equation}
Note that the condition~\eqref{eq:r3gammaconstraint} is equivalent to $r_{3}+s_{3}\ge0$, which is the condition that guarantees $XX^{\dag}=\Choi[\map{F}]$.
The background colors have been added to remind the reader how to extract Kraus operators from a Choi matrix, which are 
\begin{equation}
F_{0}=
\left(\begin{array}{cc}
\cellcolor{red!10}\sqrt{\frac{1+r_3}{1+s_3}}&\cellcolor{orange!10}0\\
\cellcolor{red!10}0&\cellcolor{orange!10}\sqrt{\frac{(1-\gamma)(1+s_3)}{1+r_3}}\end{array}\right)
,\qquad
F_{1}=
\left(\begin{array}{cc}
\cellcolor{yellow!10}0&\cellcolor{green!10}\phantom{x}0\phantom{x}\\
\cellcolor{yellow!10}\sqrt{\frac{\gamma(1-r_3)}{1+s_3}}&\cellcolor{green!10}\phantom{x}0\phantom{x}\end{array}\right)
,\qquad
F_{2}=
\left(\begin{array}{cc}
\cellcolor{blue!10}\phantom{x}0\phantom{x} & \cellcolor{violet!10}0 \\
\cellcolor{blue!10}\phantom{x}0\phantom{x} & \cellcolor{violet!10}\sqrt{\frac{\gamma(r_3+s_3)}{1+r_3}}
\end{array}\right)
\end{equation} 
with the corresponding colors indicating how they are obtained from $X$ in~\eqref{eq:XforBayesmapr3}. 

From these Kraus operators, we see that the Bayesian inverse $\map{F}$ is a combination of a bit-flipped amplitude-damping channel combined with a dephasing channel~\cite{NiCh11,JROCAG17}. In particular, the Kraus operators are of the form 
\be
F_{0}=\begin{pmatrix}\sqrt{1-\kappa}&0\\0&\sqrt{1-\lambda}\end{pmatrix}=\frac{\sqrt{1-\kappa}}{2} (\mathds{1}_{2}+\sigma_{3})+\frac{\sqrt{1-\lambda}}{2} (\mathds{1}_{2}-\sigma_{3})\, ,
\ee
\be
F_{1}=\begin{pmatrix}0&0\\\sqrt{\kappa}&0\end{pmatrix}=\frac{\sqrt{\kappa}}{2}(\sigma_{1}-i\sigma_{2})
\, , \quad\text{ and }\quad
F_{2}=\begin{pmatrix}0&0\\0&\sqrt{\lambda}\end{pmatrix}=\frac{\sqrt{\lambda}}{2}(\mathds{1}_{2}-\sigma_{3})\, , 
\ee
where
\be
\kappa=\frac{\gamma(1-r_{3})}{1+s_{3}}=\frac{\gamma(1-r_{3})}{1+r_{3}+\gamma(1-r_{3})}
\quad\text{ and }\quad
\lambda=\frac{\gamma(r_{3}+s_{3})}{1+r_3}=\frac{\gamma(2r_{3}+\gamma(1-r_{3}))}{1+r_3}\, ,
\ee
which both take values between $0$ and $1$ by the assumptions that $\gamma\in(0,1)$,  $r_{3}\in(-1,1)$, and the constraint~\eqref{eq:r3gammaconstraint}.  

We now proceed to analyze the two-time expectation values of the amplitude damping process $(\map{E},\rho)$ and its Bayesian inverse $(\map{F},\map{E}(\rho))$ for the prior density matrix $\rho$ from~\eqref{eq:priorrhoADC}. The measurements will be of arbitrary Pauli observables and we will explicitly show that forward-time expectation values agree with the backward-time expectation values, despite the fact that $\map{E}$ is neither a unitary nor a unital quantum channel. 
For reference, $\map{E}\star\rho$, for the general $\rho$ from~\eqref{eq:generalinitialqubitstate}, is given by 
\be
\map{E}\star\rho=\frac{1}{2}\big\{\rho\otimes\mathds{1}_{2},\Jamiol[\map{E}]\big\} 
=\frac{1}{4}
\begin{pmatrix}
2(1+r_3)&\sqrt{\gamma^{\perp}}\overline{r_{12}}&(1+\gamma)\overline{r_{12}}&0 \\
\sqrt{\gamma^{\perp}}r_{12}&0&2\sqrt{\gamma^{\perp}}&\gamma^{\perp}\overline{r_{12}}\\
(1+\gamma)r_{12}&2\sqrt{\gamma^{\perp}}&2\gamma(1-r_3)&\sqrt{\gamma^{\perp}}\overline{r_{12}}\\
0&\gamma^{\perp}r_{12}&\sqrt{\gamma^{\perp}}r_{12}&2\gamma^{\perp}(1-r_3)
\end{pmatrix},
\ee
where 
\begin{equation}
\gamma^{\perp}:=1-\gamma,\quad
r_{12}:=r_{1}+ir_{2},\quad\text{ and }\quad
\overline{r_{12}}=r_{1}-ir_{2}\, .
\end{equation}

Setting $r_{1}=r_{2}=0$ to analyze the special case where the input matrix is diagonal, the state over time $\map{E}\star\rho$ associated with $(\map{E},\rho)$ is given by
\begin{equation}
\map{E}\star\rho=
\frac{1}{2}
\begin{pmatrix}
1+r_{3}&0&0&0\\
0&0&\sqrt{\gamma^{\perp}}&0\\
0&\sqrt{\gamma^{\perp}}&\gamma(1-r_3)&0\\
0&0&0&\gamma^{\perp}(1-r_3)
\end{pmatrix}\, .
\end{equation}
By Lemma~\ref{MTXS45739}, the two-time expectation values of the Pauli observables for the forward process $(\map{E},\rho)$ are 
\begin{equation}
\label{eq:ADCTTEVsimplified}
\<\sigma_{\alpha},\sigma_{\beta}\>=\Tr\left[(\map{E}\star\rho)(\sigma_{\alpha}\otimes\sigma_{\beta})\right]
\end{equation}
for all $\alpha,\beta\in\{0,1,2,3\}$. These values are displayed as a matrix in Table~\ref{tab:ADCTTEVFPB}. 

Meanwhile, the quantum state over time associated with the Bayesian inverse process $(\map{F},\map{E}(\rho))$ is given by
\begin{equation}
\map{F}\star\map{E}(\rho)=
\frac{1}{2}
\begin{pmatrix}
1+r_3&0&0&0\\
0&\gamma(1-r_3)&\sqrt{\gamma^{\perp}}&0\\
0&\sqrt{\gamma^{\perp}}&0&0\\
0&0&0&\gamma^{\perp}(1-r_3)
\end{pmatrix},
\end{equation}
where we have used the Jamio{\l}kowski matrix~\eqref{eq:ChoiBayesmapr3}  
and the predicted state $\map{E}(\rho)$ from~\eqref{eq:predictionrhoADC}. Note that 
$\map{E}\star\rho=\mathcal{S}\big(\map{F}\star\map{E}(\rho)\big)$
(where $\mathcal{S}$ is the swap map), in accordance with the definition of the Bayesian inverse $\map{F}$. Therefore, when writing the two-time expectation values $\<\sigma_{\beta},\sigma_{\alpha}\>$ for the basis of Pauli observables $(\sigma_{0},\sigma_{1},\sigma_{2},\sigma_{3})$ associated with the reverse process $(\map{F},\map{E}(\rho))$ as a matrix $(b_{\beta\alpha})$, we expect this matrix to be the transpose of the matrix $(a_{\alpha\beta})$ of expectation values $\<\sigma_{\alpha},\sigma_{\beta}\>$ associated with the process $(\map{E},\rho)$, i.e., $b_{\beta\alpha}=a_{\alpha\beta}$ for all $\alpha,\beta$. This is shown explicitly in Table~\ref{tab:ADCBayesTTEVFPB}.

\begin{figure}
\includegraphics[width=5.0cm]{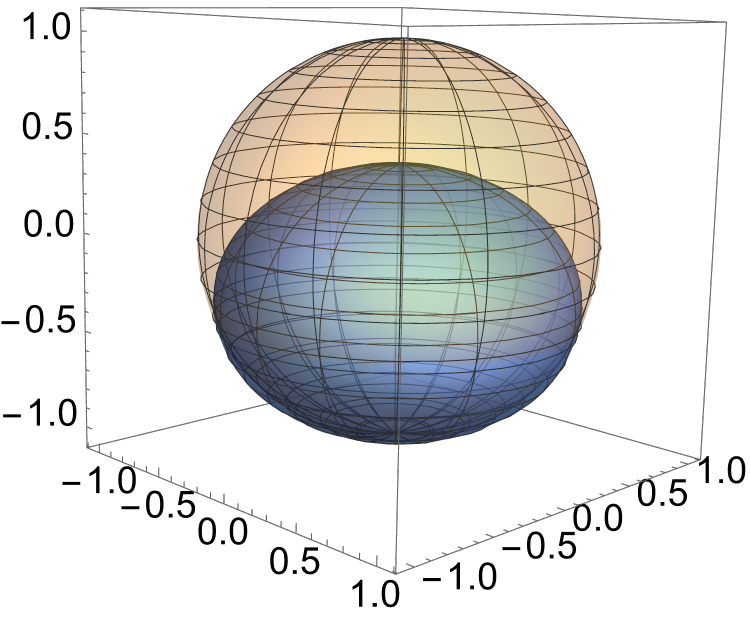}
\hspace{10mm}
\includegraphics[width=5.0cm]{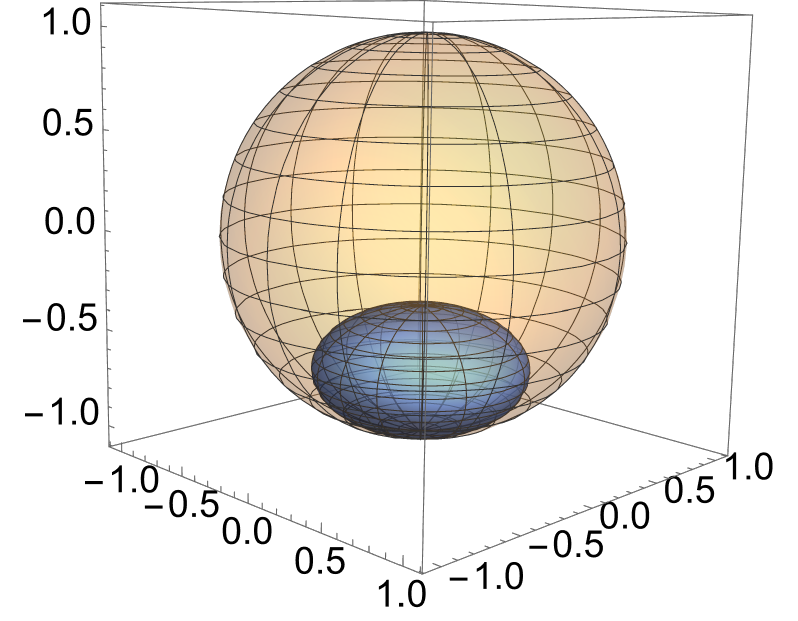}
\caption{Visualizations of the image of the Bloch ball after one application of the Bayesian inverse $\map{F}$ of the amplitude-damping channel. 
The original Bloch ball is in translucent orange, while the image after one application is shown in blue. 
On the left, $r_3=-0.5$ and $\gamma=0.15$ so that $r_{3}<\frac{\gamma}{\gamma-2}$, which guarantees the map $\map{F}$ is not completely positive. 
Indeed, the image of the Bloch ball on the left partly escapes the original Bloch ball, illustrating that $\map{F}$ is not even positive. 
Meanwhile, the right shows the analogous situation for $r_{3}=-0.5$ and $\gamma=0.7$, which satisfies $r_{3}>\frac{\gamma}{\gamma-2}$ so that $\map{F}$ is completely positive. 
It is visible that the image of the Bloch ball remains inside the original Bloch ball.}
\label{fig:ADCBayesNotPositive}
\end{figure}

We now analyze the same amplitude-damping channel but with an initial diagonal density matrix $\rho$ for which $(\map{E},\rho)$ does not admit a (completely positive) Bayesian inverse. This can occur when $r_{3}<0$. More precisely, $\map{F}$ as defined in~\eqref{eq:Binvgeneralformula} is a quantum channel if and only if $r_{3}$ satisfies~\eqref{eq:r3gammaconstraint}. 
This means that $(\map{E},\rho)$ is not Bayesian invertible for all input density matrices $\rho$ and amplitude-damping parameters $\gamma$. Nevertheless, we can examine what the map $\map{F}$, as defined in~\eqref{eq:Binvgeneralformula}, does to the Bloch ball despite it not being a quantum channel. Figure~\ref{fig:ADCBayesNotPositive} shows that the map $\map{F}$ might not even be positive as it can take the Bloch ball outside of the Bloch ball. 

Our analysis thus far is summarized by the following theorem.

\bt
\label{thm:ADCsymmetry}
Let $\map{E}:\Alg{A}\to\Alg{B}$ be the amplitude-damping channel on a single qubit with damping parameter $\gamma\in(0,1)$ (so that $\Alg{A}=\Alg{B}=\matr_{2}$), let $\rho=\frac{1}{2}(\mathds{1}+r_{3}\sigma_{3})$ be an initial state with $r_{3}\in(-1,1)$, and let $\mathscr{S}\subset\mathbf{Obs}(A)\times\mathbf{Obs}(B)$ be the subset of all pairs of qubit Pauli operators. 
Then $(\map{E},\rho)$ has a $\mathscr{S}$-operational inverse if and only if $r_{3}\ge\frac{\gamma}{\gamma-2}$, in which case the operational inverse equals the  Bayesian inverse and is given explicitly by~\eqref{eq:BayesianinverseADC}. 
\et

We now contrast the Bayesian inverse $\map{F}$ to the Petz recovery map 
$\map{R}=\Ad_{\rho^{1/2}}\circ\map{E}^*\circ\Ad_{\map{E}(\rho)^{-1/2}}$. The Petz recovery map has a Kraus representation given by
\begin{equation}
R_{0}=
\rho^{1/2}E_{0}^{\dag}\map{E}(\rho)^{-1/2}=\begin{pmatrix}\sqrt{\frac{1+r_3}{1+s_3}}&0\\0&1\end{pmatrix}
\quad\text{ and }\quad
R_{1}=\rho^{1/2}E_{1}^{\dag}\map{E}(\rho)^{-1/2}=\begin{pmatrix}0&0\\\sqrt{\frac{\gamma(1-r_3)}{1+s_3}}&0\end{pmatrix}\, .
\end{equation}
Its Choi and Jamio{\l}kowski matrices are given by 
\begin{equation}
\label{eq:JamiolPetzr3}
\Choi[\map{R}]
=\begin{pmatrix}
\frac{1+r_3}{1+s_3}&0&0&\sqrt{\frac{1+r_3}{1+s_3}}\\
0&\frac{\gamma(1-r_3)}{1+s_3}&0&0\\
0&0&0&0\\
\sqrt{\frac{1+r_3}{1+s_3}}&0&0&1
\end{pmatrix}
\quad\text{ and }\quad
\Jamiol[\map{R}]=
\begin{pmatrix}
\frac{1+r_3}{1+s_{3}}&0&0&0\\
0&\frac{\gamma(1-r_3)}{1+s_{3}}&\sqrt{\frac{1+r_3}{1+s_{3}}}&0\\
0&\sqrt{\frac{1+r_3}{1+s_{3}}}&0&0\\
0&0&0&1
\end{pmatrix}\, ,
\end{equation}
respectively. 
Note that $\Choi[\map{R}]$ is positive for \emph{all} values of $r_{3}\in(-1,1)$ and $\gamma\in(0,1)$. 

Using Lemma~\ref{MTXS45739} to simplify the computations of two-time expectation values,  
the associated state over time is
\begin{equation}
\map{R}\star\map{E}(\rho)=
\frac{1}{2}
\begin{pmatrix}
1+r_3&0&0&0\\
0&\gamma(1-r_3)&\sqrt{\frac{1+r_3}{1+s_{3}}}&0\\
0&\sqrt{\frac{1+r_3}{1+s_{3}}}&0&0\\
0&0&0&\gamma^{\perp}(1-r_3)
\end{pmatrix},
\end{equation}
where $\gamma^{\perp}:=1-\gamma$. 
The associated two-time expectation values are shown in Table~\ref{tab:ADCPetzTTEVFPB}. Although these expectation values can be experimentally verified using the Petz recovery map $\map{R}$, one can see that these do \emph{not} in general agree with the two-time expectation values of the Bayesian inverse $\map{F}$. In particular, they do not reproduce the (transpose of) the original two-time expectation values for the forward process given by the amplitude-damping channel $\map{E}$. As such, the Petz recovery map does not admit a time symmetry in the sense advocated for two-time expectation values of light-touch observables.

\begin{table}
\begin{tabular}{c|cccc}
&$\sigma_{0}$&$\sigma_{1}$&$\sigma_{2}$&$\sigma_{3}$\\
\hline
$\sigma_{0}$ & $1$ & $0$ & $0$ & $r_3$ \\
$\sigma_{1}$ & $0$ & $\sqrt{\frac{1+r_3}{1+s_3}}$ & $0$ & $0$ \\
$\sigma_{2}$ & $0$ & $0$ & $\sqrt{\frac{1+r_3}{1+s_3}}$ & $0$ \\
$\sigma_{3}$ & $s_3$ & $0$ & $0$ & $1-\gamma(1-r_3)$\\
\end{tabular}
\hspace{10mm}
\begin{tabular}{c|cccc}
&$\sigma_{0}$&$\sigma_{1}$&$\sigma_{2}$&$\sigma_{3}$\\
\hline
$\sigma_{0}$ & $1$ & $0$ & $0$ & $r_3$ \\
$\sigma_{1}$ & $0$ & $\sqrt{1-\gamma}$ & $0$ & $0$ \\
$\sigma_{2}$ & $0$ & $0$ & $\sqrt{1-\gamma}$ & $0$ \\
$\sigma_{3}$ & $s_3$ & $0$ & $0$ & $1-\gamma(1-r_3)$\\
\end{tabular}
\caption{The two-time expectation values for the Petz recovery map $\map{R}$ of the amplitude-damping channel $\map{E}$ with initial density matrix $\rho$ from~\eqref{eq:priorrhoADC} is shown on the left. Each row corresponds to a fixed measurement for Bob (first measurement) and each column corresponds to a fixed measurement for Alice (second measurement). Here, $s_3=r_3+\gamma(1-r_3)$. In this way, the table specifies a matrix whose $(\beta,\alpha)$ entry is $\<\sigma_{\beta},\sigma_{\alpha}\>$ using the Petz recovery map process $(\map{R},\map{E}(\rho))$. Unlike the case of the Bayesian inverse from Table~\ref{tab:ADCBayesTTEVFPB}, which is reproduced on the right for convenience, this matrix is \emph{not} the transpose of the matrix from Table~\ref{tab:ADCTTEVFPB}. The main differences are the two-time expectation values $\<\sigma_{j},\sigma_{j}\>$ for $j\in\{1,2\}$.}
\label{tab:ADCPetzTTEVFPB}
\end{table}

One might object that it is unnatural to use the expression $\map{R}\star\map{E}(\rho)=\frac{1}{2}\big\{\map{E}(\rho)\otimes\mathds{1}_{2},\Jamiol[\map{R}]\big\}$ and that one should instead use what is known as the Leifer--Spekkens spatiotemporal product $\star_{\text{LS}}$~\cite{LeSp13,HHPBS17,FuPa22a}, which has the Petz recovery map as its Bayesian inverse~\cite{FuPa22a}. The Leifer--Spekkens spatiotemporal product $\star_{\text{LS}}$ is given by 
\begin{equation}
\map{E}\star_{\mathrm{LS}}\rho=(\sqrt{\rho}\otimes\mathds{1}_{2})\Jamiol[\map{E}](\sqrt{\rho}\otimes\mathds{1}_{2})\, ,
\end{equation}
so that
\begin{equation}
\map{R}\star_{\mathrm{LS}}\map{E}(\rho)=(\sqrt{\map{E}(\rho)}\otimes\mathds{1}_{2})\Jamiol[\map{R}](\sqrt{\map{E}(\rho)}\otimes\mathds{1}_{2})\, .
\end{equation}
However, the formal mathematical expressions 
\begin{equation}
\Tr\left[\big(\map{E}\star_{\mathrm{LS}}\rho\big)(\sigma_{\alpha}\otimes\sigma_{\beta})\right]
\quad\text{ and }\quad
\Tr\left[\big(\map{R}\star_{\mathrm{LS}}\map{E}(\rho) \big)(\sigma_{\beta}\otimes\sigma_{\alpha})\right]
\end{equation}
no longer admit the operational interpretation of two-time expectation values of Pauli observables~\cite{FuPa24a} (however, see Refs.~\cite{Le06,Le07,FuPa22a} for an operational interpretation in terms of prepare-evolve-measure scenarios). We explicitly show this in Table~\ref{tab:LSvalues} by computing these two expressions and comparing them to the true values of the two-time expectation values. Although they are symmetric with respect to each other, they do not admit the same operational meaning as two-time expectation values. 

\begin{table}
(a)
\begin{tabular}{c|cccc}
&$\sigma_{0}$&$\sigma_{1}$&$\sigma_{2}$&$\sigma_{3}$\\
\hline
$\sigma_{0}$ & $1$ & $0$ & $0$ & $r_3$ \\
$\sigma_{1}$ & $0$ & $\sqrt{(1-\gamma)(1-r_3^2)}$ & $0$ & $0$ \\
$\sigma_{2}$ & $0$ & $0$ & $\sqrt{(1-\gamma)(1-r_3^2)}$ & $0$ \\
$\sigma_{3}$ & $s_3$ & $0$ & $0$ & $1-\gamma(1-r_3)$\\
\end{tabular}
%
\hfill
(b)
\begin{tabular}{c|cccc}
&$\sigma_{0}$&$\sigma_{1}$&$\sigma_{2}$&$\sigma_{3}$\\
\hline
$\sigma_{0}$ & $1$ & $0$ & $0$ & $s_3$ \\
$\sigma_{1}$ & $0$ & $\sqrt{(1-\gamma)(1-r_3^2)}$ & $0$ & $0$ \\
$\sigma_{2}$ & $0$ & $0$ & $\sqrt{(1-\gamma)(1-r_3^2)}$ & $0$ \\
$\sigma_{3}$ & $r_3$ & $0$ & $0$ & $1-\gamma(1-r_3)$\\
\end{tabular}
\caption{(a) A table of the values $\Tr\left[\big(\map{E}\star_{\mathrm{LS}}\rho\big)(\sigma_{\alpha}\otimes\sigma_{\beta})\right]$ associated with the forward process $(\map{E},\rho)$. (b) A table of the values $\Tr\left[\big(\map{R}\star_{\mathrm{LS}}\map{E}(\rho) \big)(\sigma_{\beta}\otimes\sigma_{\alpha})\right]$ associated with the Petz reverse process $(\map{R},\map{E}(\rho))$. Although these two tables are transposes of each other and therefore admit a symmetry, they do not agree with the operational interpretation of providing the forward or reverse two-time expectation values.}
\label{tab:LSvalues}
\end{table}

\begin{figure}
\includegraphics[width=5.0cm]{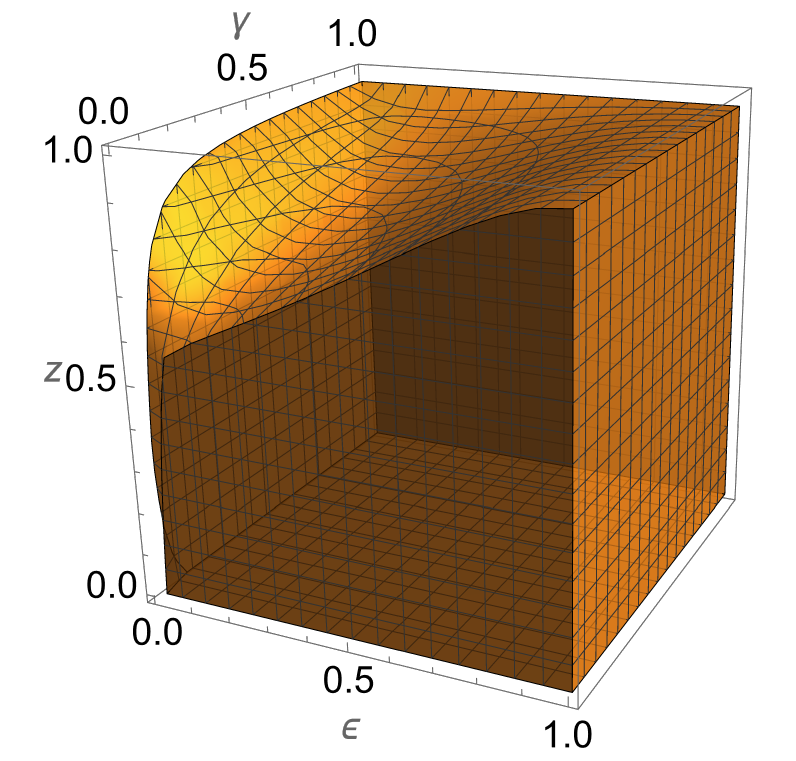}
\caption{The set of values for $\epsilon\in(0,1)$, $\gamma\in(0,1)$, and $z=r_{3}\in(0,1)$ for which the Bayesian inverse of the map $\map{E}$ given by~\eqref{eq:ADCmixCD} and state $\rho=\frac{1}{2}(\mathds{1}_{2}+r_{3}\sigma_{3})$ is a valid quantum channel.}
\label{fig:robustregion}
\end{figure}

In the remainder of this appendix, we will work out the robustness of Bayesian inverses. 
Numerical calculations suggest that there does not exist an open set of density matrices containing a density matrix of the form $\rho=\frac{1}{2}(\mathds{1}_{2}+r_{3}\sigma_{3})$ with $r_{3}$ satisfying~\eqref{eq:r3gammaconstraint}. This might seem problematic for experimental realizations because preparing quantum states to infinite precision is impossible. To remedy this, we provide a slight variation of the amplitude-damping channel by adding completely depolarizing noise. More precisely, for any $\epsilon\in[0,1]$ and $\gamma\in(0,1)$, let 
\be
\label{eq:ADCmixCD}
\map{E}=(1-\epsilon)\big(\Ad_{E_{0}}+\Ad_{E_{1}}\big)+\frac{\epsilon \mathds{1}_{2}}{2}\Tr[\;\cdot\;]
\ee
be the $\epsilon$-weighted mixture of the amplitude-damping channel with the completely depolarizing channel on a single qubit, where the Kraus operators $E_{0}$ and $E_{1}$ are as in~\eqref{eq:ADCKrausOperators}. With this modification and an initial state $\rho$ of the form $\rho=\frac{1}{2}(\mathds{1}_{2}+\vec{r}\cdot\vec{\sigma})$, the output state $\map{E}(\rho)$ is of the form 
\be
\map{E}(\rho)=\frac{1}{2}\big(\mathds{1}_{2}+(1-\epsilon)\vec{s}\cdot\vec{\sigma}\big)\, ,
\ee
where $\vec{s}$ is just as in~\eqref{eq:Erhoimages}. 
Momentarily restricting again to the case where $r_{1}=r_{2}=0$ so that the eigenbasis for $\map{E}(\rho)$ can be taken as the computational basis $|0\>,|1\>$, we can explicitly compute the Choi matrix of the Bayesian inverse $\mathcal{F}$ as a function of the parameters $\gamma\in(0,1)$, $\epsilon\in[0,1]$, and $r_{3}\in(-1,1)$ using~\eqref{eq:Binvgeneralformula}. The result is 
\be
\Choi[\map{F}]=
\begin{pmatrix}
\frac{(1-\epsilon/2)(1+r_3)}{1+(1-\epsilon)(r_{3}+\gamma(1-r_{3}))}&0&0&(1-\epsilon)\sqrt{1-\gamma}\\
0&\frac{((1-\epsilon)\gamma+\epsilon/2)(1-r_{3})}{1+(1-\epsilon)(r_{3}+\gamma(1-r_{3}))}&0&0\\
0&0&\frac{(\epsilon/2)(1+r_3)}{1-(1-\epsilon)(r_{3}+\gamma(1-r_{3}))}&0\\
(1-\epsilon)\sqrt{1-\gamma}&0&0&\frac{((1-\epsilon)(1-\gamma)+\epsilon/2)(1-r_3)}{1-(1-\epsilon)(r_{3}+\gamma(1-r_{3}))}\\
\end{pmatrix}
.
\ee
Using the subleading determinant test for positivity~\cite{Strang2022}, we find that each of the top left three diagonal entries are \emph{always} positive provided that $\epsilon>0$ and $r_{3}\in(-1,1)$. As for the final determinant, since the two middle diagonal entries are positive under these assumptions, positivity of the determinant of $\Choi[\map{F}]$ is equivalent to positivity of the $2\times 2$ block obtained by removing the second and third columns and rows of $\Choi[\map{F}]$. Thus, positivity of this determinant is equivalent to 
\be
\frac{(1-\epsilon/2)((1-\epsilon)(1-\gamma)+\epsilon/2)(1-r_{3}^2)}{1-(1-\epsilon)^2(r_{3}+\gamma(1-r_{3}))^2} > (1-\epsilon)^2(1-\gamma)\, . 
\ee
The region described by this inequality is sketched in Figure~\ref{fig:robustregion}. By construction, it is an open subset of the unit cube $(0,1)\times(0,1)\times(0,1)$ in $\R^{3}$. We now state and prove the robustness of Bayesian inverses in this setting. 

\bt
\label{thm:ADCrobustness}
Fix $\epsilon,\gamma\in(0,1)$ and let $\map{E}$ be the amplitude-damping channel mixed with the completely depolarizing  channel as in~\eqref{eq:ADCmixCD}. 
Let $\rho=\frac{1}{2}(\mathds{1}_{2}+r_{3}\sigma_{3})$ be a density matrix such that a Bayesian inverse of $(\map{E},\rho)$ exists. 
Then there exists a $\Delta>0$ such that 
$\rho+\delta h$ is a density matrix and
$(\map{E},\rho+\delta h)$ is Bayesian invertible
for all $\delta\in[0,\Delta)$ and for all $2\times 2$ traceless hermitian matrices $h$ such that $\lVert h \rVert\le 1$. 
\et

In other words, Theorem~\ref{thm:ADCrobustness} says that for every $r_{3}\in(0,1)$ such that $(\epsilon,\gamma,r_3)$ is in the region of Figure~\ref{fig:robustregion}, there exists an open neighborhood $U\subset\mathbb{D}^{3}$, where $\mathbb{D}^{3}$ is the unit disc in $\R^{3}$, containing $(0,0,r_{3})$ and for which 
\be
\left(\map{E},\frac{1}{2}(\mathds{1}_{2}+\vec{r}\cdot\vec{\sigma})\right)
\ee
is Bayesian invertible for all $\vec{r}\in U$. 
For experimental purposes, this means that by working with the channel $\map{E}$ as in~\eqref{eq:ADCmixCD}, one can, in principle, prepare an imperfect state $\rho$ (i.e., within some error bounds) for which $(\map{E},\rho)$ is still Bayesian invertible. 
The following proof will involve some familiarity with operator norms and analysis as covered in Refs.~\cite{Fo07,Rudin76,Rudin87}, for example. 

\begin{proof}[Proof of Theorem~\ref{thm:ADCrobustness}]
Using results and techniques from Ref.~\cite{FuPa22a} (specifically Lemma 2 in Appendix A in Ref.~\cite{FuPa22a}), Bayes' rule requires 
\be
\big\{\mathds{1}_{2}\otimes\rho,\Choi[\map{E}^*]\big\}=\big\{\map{E}(\rho)^{T}\otimes\mathds{1}_{2},\Choi[\map{F}]\big\}\, ,
\ee
where the transpose ${}^{T}$ is taken with respect to the standard basis.  
Solving for $\Choi[\map{F}]$ using the integral solution of the Sylvester equation (cf.\ Refs.~\cite{Pe71,BhRo97,Hi08,LCD24}) yields
\be
\Choi[\map{F}]=\int_{0}^{\infty}\big(e^{- s \map{E}(\rho)^{T}}\otimes\mathds{1}_{2}\big)\big\{\mathds{1}_{2}\otimes\rho,\Choi[\map{E}^*]\big\}\big(e^{- s \map{E}(\rho)^{T}}\otimes\mathds{1}_{2}\big)\,ds\, .
\ee
Now replace $\rho$ with $\rho+\delta h$, where $h$ is some fixed traceless hermitian matrix, $\delta\in[-\Delta',\Delta']$, and $\Delta'\in(0,1)$ is small enough so that $\map{E}(\rho)^{T}+\delta\map{E}(h)^{T}$ is a strictly positive density matrix for all $\delta\in[-\Delta',\Delta']$. Note that such a $\Delta'$ exists as $\map{E}(\rho)^{T}$ is full rank and $\map{E}(h)^{T}$ is traceless and hermitian (and since the real tangent space at a full rank density matrix is isomorphic to the real vector space of traceless hermitian matrices~\cite{BeZy06,Petz96,SASDS23}). 
Set
\be
\lambda_{0}:=\min_{\delta\in[-\Delta',\Delta']}\lambda_{\min}\Big(\map{E}(\rho)^{T}+\delta \map{E}(h)^{T}\Big)\, , 
\ee
where $\lambda_{\min}(A)$ denotes the minimum (necessarily strictly positive) eigenvalue of a strictly positive matrix $A$. By assumption on $\Delta'$, it is necessarily the case that $\lambda_{0}>0$. Now, for each $\delta\in[-\Delta',\Delta']$, set 
\be
X_{\delta}(s):=\big(e^{- s \map{E}(\rho+\delta h)^{T}}\otimes\mathds{1}_{2}\big)\big\{\mathds{1}_{2}\otimes(\rho+\delta h),\Choi[\map{E}^*]\big\}\big(e^{- s \map{E}(\rho+\delta h)^{T}}\otimes\mathds{1}_{2}\big)\, .
\ee
Note that the operator $Y_{\delta}:=\int_{0}^{\infty}X_{\delta}(s)\,ds$ is well-defined for all $\delta\in[-\Delta',\Delta]$ and satisfies $Y_{0}=\Choi[\map{F}]$, again by the assumption on $\Delta'$. 
The goal in proving the theorem is to show $\lim_{\delta\to0}Y_{\delta}= Y_{0}$, which would prove that the Choi matrix $\Choi[\map{F}]$ of the Bayesian inverse $\map{F}$ of $(\map{E},\rho)$ is continuous at $\rho$~\cite{Fo07,Rudin76}.  
Now let $g:[0,\infty)\to\R$ be the function defined by 
\be
g(s):=2\big\lVert\Choi[\map{E}^*]\big\rVert \big(\lVert\rho\rVert+\lVert h\rVert\big)e^{-2\lambda_{0}s}.
\ee
Note that $g$ is a non-negative $L_{1}$-integrable function with respect to the Lebesgue measure on $[0,\infty)$, since its integral is 
\be
\int_{0}^{\infty}g(s)\,ds=\frac{\lVert\Choi[\map{E}^*]\rVert (\lVert\rho\rVert+\lVert h\rVert)}{\lambda_{0}}.
\ee
For any pair of unit vectors $|\psi\>\in\C^{n}$ and $|\phi\>\in\C^{m}$, let $f_{\delta}:[0,\infty)\to\R$ be the function defined by 
\be
f_{\delta}(s):=\<\psi,\phi|X_{\delta}(s)|\psi,\phi\>\, ,
\ee
which is real-valued as the operator $X_{\delta}(s)$ is hermitian for all $s\in[0,\infty)$ and $\delta\in[-\Delta',\Delta']$. 
One then has
\begin{align}
|f_{\delta}(s)|&\le \lVert X_{\delta}(s)\rVert \nonumber \\
&\le \Big\lVert e^{-s(\map{E}(\rho)^{T}+\delta\map{E}(h)^{T})}\otimes\mathds{1}_{2}\Big\rVert^{2}\bigg(\Big\lVert\big\{\mathds{1}_{2}\otimes\rho,\Choi[\map{E}^*]\big\}\Big\rVert+|\delta|\Big\lVert\big\{\mathds{1}_{2}\otimes h,\Choi[\map{E}^*]\big\}\Big\rVert\bigg) \nonumber \\
&\le 2 \Big\lVert e^{-s(\map{E}(\rho)^{T}+\delta\map{E}(h)^{T})}\Big\rVert^2\big\lVert\Choi[\map{E}^*]\big\rVert\Big(\lVert\rho\rVert+|\delta|\,\lVert h\rVert\Big) \nonumber\\
&\le g(s)\, .
\label{eq:lemmainequalityproof}
\end{align}
The first inequality follows from the property $|\<\eta|A|\eta\>|\le\lVert A\rVert\<\eta|\eta\>=\lVert A\rVert$ for all unit vectors $|\eta\>$ and for all operators $A$. 
The second inequality follows from the properties $\lVert AB\rVert\le\lVert A\rVert \lVert B\rVert$ and $\lVert A+\delta B\rVert\le\lVert A\rVert+|\delta|\lVert B\rVert$ for all operators $A,B$ and real numbers $\delta$. 
The third inequality follows from the same properties as above together with $\lVert A\otimes\mathds{1}\rVert=\lVert A\rVert$ for all operators $A$.
The final inequality follows from the assumption that $\Delta'<1$, the inequality $\lVert e^{-s A}\rVert\le e^{-s \lVert A\rVert}$, the relationship between the operator norm of a matrix and its singular values, and the definition of $\lambda_{0}$. 
The inequality~\eqref{eq:lemmainequalityproof} says that $|f_{\delta}(s)|\le g(s)$ for all $\delta\in[-\Delta,\Delta']$ and $s\in[0,\infty)$. Hence, 
\be
\lim_{\delta\to0}\<\psi,\phi| Y_{\delta}|\psi,\phi\>=\lim_{\delta\to0}\left\<\psi,\phi\left|\int_{0}^{\infty}X_{\delta}(s)\,ds\right|\psi,\phi\right\>
=\lim_{\delta\to0}\int_{0}^{\infty}f_{\delta}(s)\,ds
=\int_{0}^{\infty}f_{0}(s)\,ds
=\big\<\psi,\phi\big|\Choi[\map{F}]\big|\psi,\phi\big\>\, ,
\ee
where Lebesgue's dominated convergence theorem~\cite{Rudin87,Fo07} was used in the third equality. Since this is true for all unit vectors $|\psi\>\in\C^{n}$ and $|\phi\>\in\C^{m}$, ~\cite[Lemma A.2]{FuPa24a} then implies
\be
\lim_{\delta\to0}Y_{\delta}=\Choi[\map{F}]\, ,
\ee
as needed. Therefore, by continuity, there exists a $\Delta''\in(0,\Delta']$ such that $(\map{E},\rho+\delta h)$ is Bayesian invertible for all $\delta\in[-\Delta'',\Delta'']$. By restricting $h$ to lie on the unit sphere in the space of real traceless matrices, compactness of this unit sphere implies that there exists a $\Delta>0$ such that $\rho+\delta h$ is a density matrix and
$(\map{E},\rho+\delta h)$ is Bayesian invertible
for all $\delta\in[0,\Delta)$ and for all $2\times 2$ traceless hermitian matrices $h$ such that $\lVert h \rVert\le 1$. This concludes the proof.
\end{proof}



\end{document}